\documentclass[twocolumn,notitlepage,showpacs,prl,superscriptaddress]{revtex4-2}
\usepackage{amsmath}
\usepackage{epsf}
\usepackage{graphicx}
\usepackage{dcolumn}
\usepackage{bm}

\newcommand{\beq}{\begin{equation}}
\newcommand{\eeq}{\end{equation}}
\newcommand{\be}{\begin{equation}}
\newcommand{\ee}{\end{equation}}
\newcommand{\beqa}{\begin{eqnarray}}
\newcommand{\eeqa}{\end{eqnarray}}
\newcommand{\bea}{\begin{align}}
\newcommand{\eea}{\end{align}}
\usepackage{amssymb}
\usepackage{array}
\usepackage{amsmath}
\usepackage{graphicx}\usepackage{graphics}
\usepackage{dcolumn}
\usepackage{bm}\usepackage{varioref}

\usepackage{amsmath}
\usepackage{epsf}
\usepackage{graphicx}

\usepackage{subfigure}
\usepackage{dcolumn}
\usepackage{bm}

\usepackage{amssymb}
\usepackage{array}
\usepackage{varioref}
\usepackage{wrapfig}
\usepackage{etoolbox}
\usepackage{color}

\makeatletter
\newcommand*{\balancecolsandclearpage}{%
  \close@column@grid
  \clearpage
  \twocolumngrid
}
\makeatother

\begin{document}

\title{
 Slow relaxation and aging in the model of randomly connected cycles  network }
\author{S. Reich, S. Maoz, Y. Kaplan, H. Rappeport, N.Q. Balaban, and  O. Agam}
\affiliation{The Racah Institute of Physics, The Hebrew University of Jerusalem, 9190401, Israel}

\date{\today}
\begin{abstract}

We propose a statistical model of a large random network with high connectivity in order to describe the behavior of {\it E.\,coli} cells after exposure to  acute stress. The building blocks of this network are feedback cycles typical of the genetic and metabolic networks of a cell. Each node on the cycles is a spin degree of freedom representing a component in the cell's network that can be in one of two states - active or inactive. The cycles are interconnected by regulation or by the exchange of metabolites. Stress is realized by an external magnetic field that drives the nodes into an inactive state, and the time the magnetization passes zero value for the first time represents the first division event of the cell after the stress period. The numerical and analytical solutions for this first passage problem reproduce the aging dynamics observed in the experimental data. 
\end{abstract}
\pacs{\\87.17.-d ~~~~Cell processes\\
05.40.-a ~~~~Fluctuation phenomena, random processes, noise, and Brownian motion\\
75.10.Nr ~~~Spin-glass and other random models}

 \maketitle

\section{Introduction}

The behavior of the cellular network of molecular components in living cells, from a physicist's viewpoint, is an intriguing problem \cite{parisi1993statistical}. On  one hand, a cell can be viewed as a large computer, evolved by evolution over billions of years, and programmed to deal with various environmental changes. In particular, the cell's network typically has  built-in adaptation mechanisms that allows it to withstand moderate starvation conditions by taking the cell into a new state that preserves its vitality. On the other hand, the cell is also a complex physical system with intricate  interactions and collective behavior amenable to some sort of statistical description.  

Most experimental and theoretical studies of cells exposed to stress conditions are focused on the regulatory regime. These aim to decipher the various layers of the cellular network (metabolism, gene regulation, etc.) by disentangling its various pathways, either by focusing on specific small modules \cite{ozbudak2004multistability} or in similarity to the deciphering of a computer's blueprint \cite{milo2002network} . However, this reductionist approach is limited because the cell's cellular network is incredibly complicated and strongly intertwined. 

Nevertheless, several approaches have been proposed to model the cell's network using statistical models \cite{kauffman2003random,li2013generic,edwards2000combinatorial}, and using similar approaches as the ones used in neural network theory \cite{stern2014dynamics}. The observation that exposing cells to unforeseen challenges (by genetic engineering) \cite{Stern07,Braun15} accentuates the statistical nature of their behavior as complex systems \cite{schreier2017exploratory} suggests an alternative
approach to advance our understanding of these complicated systems. In a recent work \cite{kaplan2021observation}, our goal was  to study the cell in a regime beyond its ability to respond to environmental conditions, under perturbations that drive the cell to the edge of death but do not kill, and where it behaves as a random complex system. In this regime, the cell's dynamics are dictated by the global network architecture, which is essentially random. Hence, it can be described by a relatively simple statistical model. Such a  model can be used as a starting point for developing an increasingly improved description of the cellular dynamics. The premise of this approach is that the statistical behavior of the cell holds valuable information about the functionality and the architecture of the cellular network.  

In particular, recent experimental data \cite{kaplan2021observation} show that the recovery time of {\it E.\,coli} cells, after their growth was arrested by acute stress, exhibits distinctive statistical features typical of complex physical systems. Acute stress can be achieved, for instance, by adding serine hydroxamate (SHX) to exponentially growing cultures. SHX induces an artificial starvation for the amino acid serine and results in  growth arrest \cite{tosa1971effect}. In the experiment, cells are exposed to SHX for a duration $t_w$ (waiting time), after which it is washed out, and the stress period ends. Next, the single-cell lag time, i.e. the time between the end of the stress and the first post-stress division, is recorded, and the recovery time distribution is measured from a large ensemble of cells.

\begin{figure}
\hspace{0.0cm}\includegraphics[width=0.9\columnwidth]{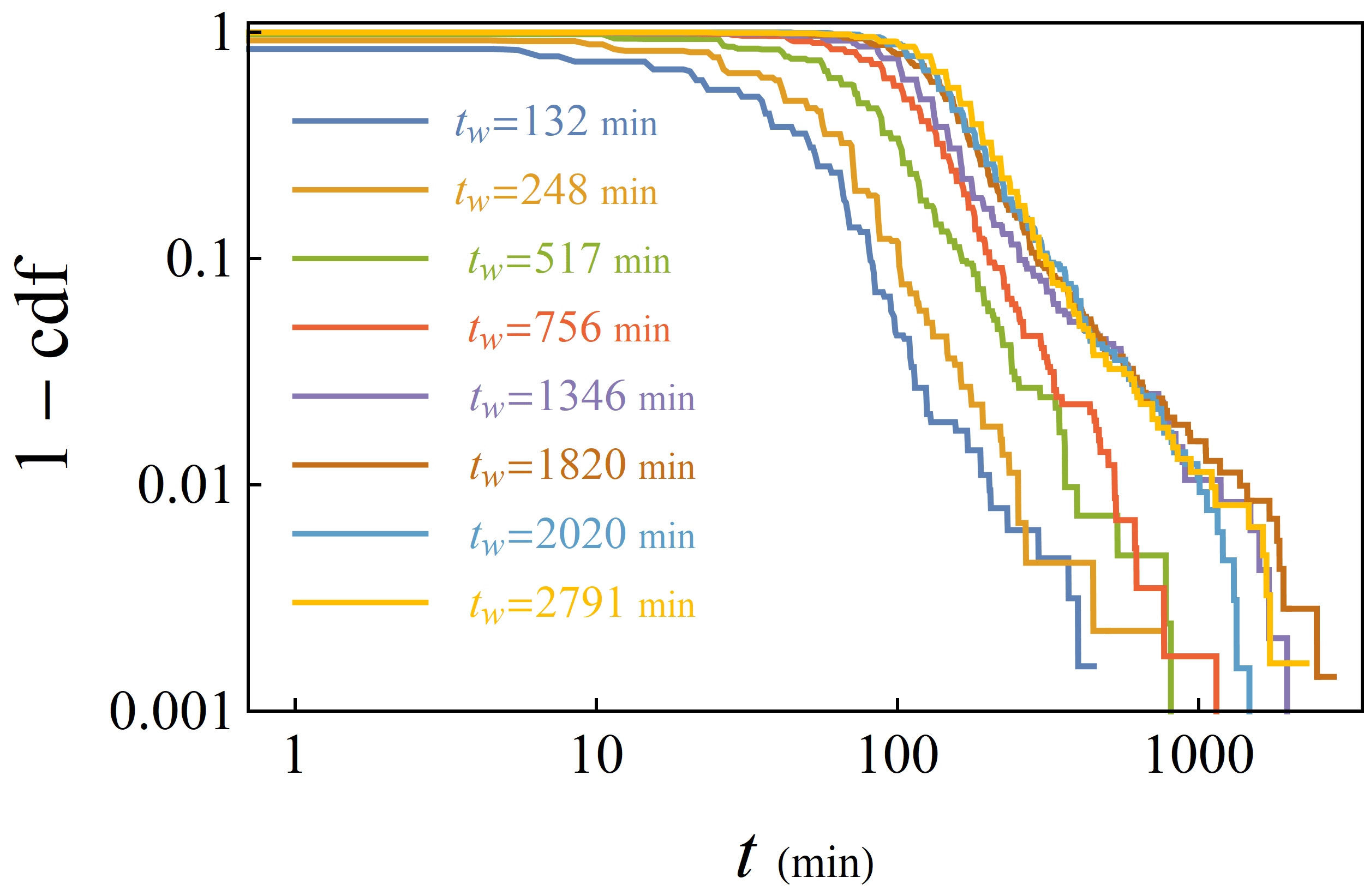} \hspace{0.9cm}
\caption{
The experimental result of 1$-$cdf associated with the distribution of first division time of cells after various starvation periods $t_w$. (Adapted from \cite{kaplan2021observation})} 
\label{Fig1}
\end{figure}

The experimental data is conveniently presented by plotting
1$-$cdf (cdf stands for cumulative distribution function), i.e., the fraction of bacteria still in the lag phase. A log-log plot of 1$-$cdf  as a function of time, for various values of the waiting time, $t_w$, is presented in Fig.~\ref{Fig1} \cite{kaplan2021observation}. This figure highlights three distinctive features of the system dynamics: First, the lag-time distribution depends on the waiting time; namely, it features a memory effect. Second, the tail of the distribution
function exhibits a slow (approximately) power-law decay.
Third, the distribution function saturates at long waiting times. The same features are common to finite physical systems that exhibit aging \cite{Ritort2003}, such as amorphous polymers \cite{Struik78,hodge1995physical}, stretched DNA \cite{Hwa2003}, paper crumpling \cite{Matan2002, Lahini2017}, colloidal solutions \cite{Ghofraniha2007a,Ghofraniha2007b}, spin glasses \cite{GlassBook,Lundgren83,cugliandolo1993analytical,vincent1997slow}, supercooled liquids \cite{Debenedetti2001}, and coulomb glasses \cite{Vaknin2001}.

We reiterate that aging behavior only appears when the cell
experiences acute stress. It is not manifested when the stress
is weak or gradual (e.g., when starvation is by natural depletion
of nutrients or by gradual application of SHX). In this case,
the cells adapt by upregulating intracellular production,
increasing import, or switching to a different metabolism that
prepares cells for survival and regrowth when nutrients become
available. With gradual stress, 1$-$cdf is practically independent
of $t_w$ and decays exponentially in time \cite{kaplan2021observation}.

In this work we construct  and solve a minimal toy model that reproduces, qualitatively, the results shown in Fig.~\ref{Fig1}.  The general approach is similar to that of Random Matrix Theory for chaotic quantum systems \cite{MehtaBook}, where a system with complex interactions and chaotic dynamics, such as atomic nuclei, is regarded as a 'black box' with a Hamiltonian matrix drawn from a Gaussian random distribution. This approach reproduces correctly the universal statistical features of the energy spectra and eigenfunctions of quantum chaotic systems in the proper energy regime.

In the same spirit, we consider the cell to be a `black box' with unknown and complicated interactions among its components (proteins, metabolites, enzymes, RNA,  etc.). To simplify the problem, we assume the following:  (a)  Each cell component is described by a boolean variable, i.e., a  spin, accounting for only two possible states of the component: active or inactive. (b)  The time evolution is discrete, and the state of the spins at a given time is determined,  deterministically, by the state of the system at its previous time step (we use zero-temperature Glauber dynamics \cite{glauber1963time}). (c) The web of interconnections among the spins (i.e., the cell components) is entirely random.

The most natural candidate for a description of this type is the Sherrington Kickpartick model \cite{SK} with zero-temperature Glauber dynamics. This remarkably simple archetype model of glasses exhibits aging, and slow relaxation  \cite{Palmer82,Kinzel86,Kinzel87,Sibani89,Kohring91,Bouchaud92,Parisi93,Kurchan94a,Kurchan94b,Kinzel95,Yoshino97} as we require. However, being Hamiltonian, the SK model features a reciprocity property not shared by the cellular network - the coupling constants between the $i$-th spin and the $j$-th spin are symmetric, $J_{ij}=J_{ji}$. However,  living cells lack this symmetry because, for example, enzymes catalyze the synthesis of products but not vice versa.   

The asymmetric  Sherrington Kickpartick model, in which all coupling constants,  $J_{ij}$, are statistically independent, is also ruled out because it exhibits, essentially, an instantaneous decay of correlations   \cite{Opper94,Parisi98}. Thus, some elements of the cell's network architecture should be retained to obtain slow relaxation and aging from an asymmetric spin network.

Here we propose a model that highlights the metabolic cycles structure of the cell by choosing its primary constituents to be closed cycles of spins rather than single spins. The sizes of these loops span over a large range of timescales, from very short timescales to long ones such as those required for protein degradation (2 min to 50 hours in growth-arrested cells \cite{nagar2021harnessing}). There are indications that these time delays are power-law distributed.

\begin{figure}[btp]
\vspace{0.2cm}
\hspace{0.0cm}\includegraphics[width=0.9\columnwidth]{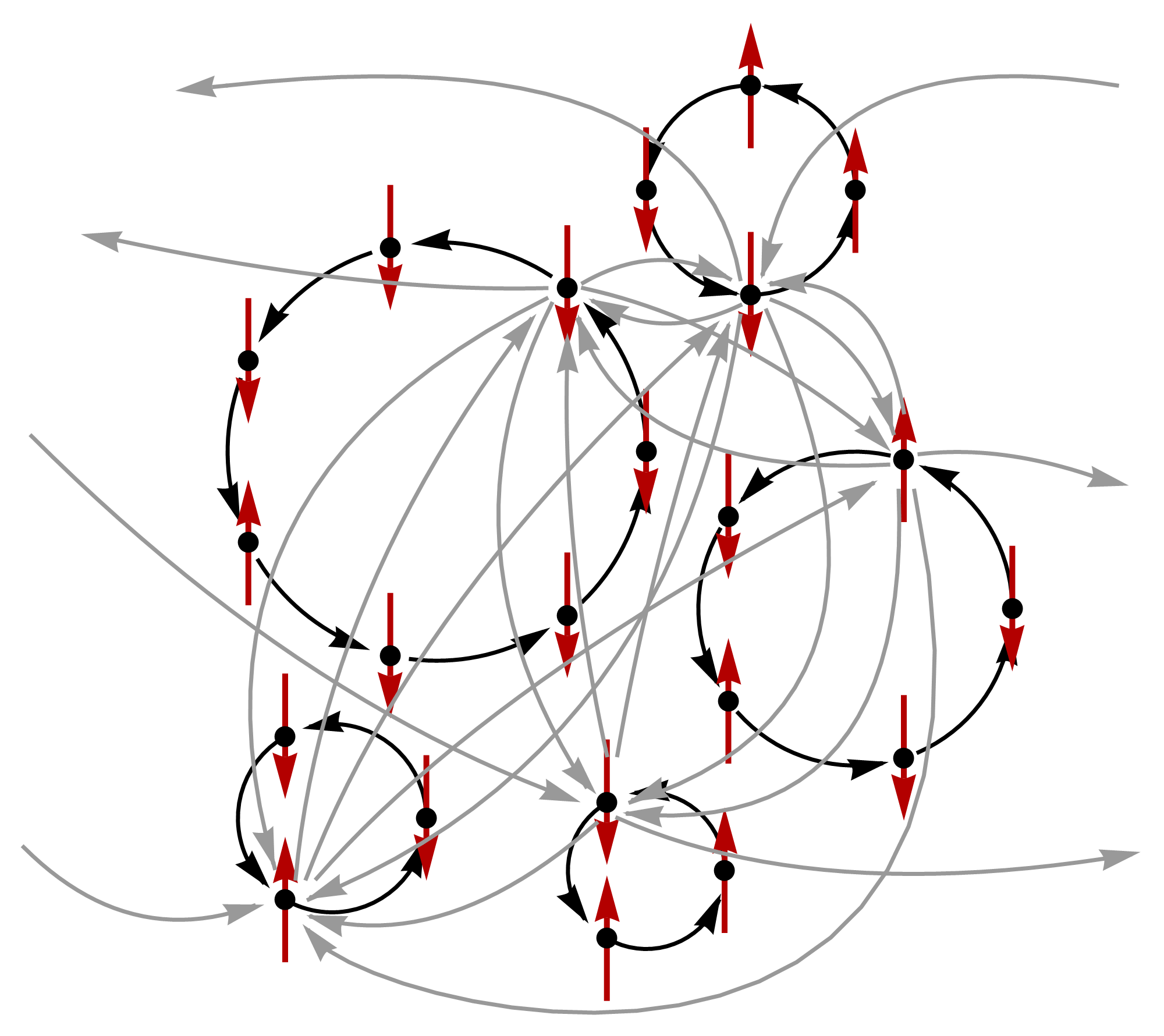}
\caption{An illustration of the Randomly Connected Cycles Network (RCCN) model. Each site in the network represents a cellular component, and any red arrow may assume two positions (up and down), denoting the corresponding component state - active or inactive. The black arrows stand for a strong coupling that forms the cycles, while gray arrows represent weak coupling between different cycles. The coupling is via a single site in each cycle.} 
\label{Fig2}
\end{figure}

Thus the toy model proposed here consists of a collection of closed spin chains (representing, for example, different metabolic cycles, or closed regulatory feedbacks) randomly interconnected via a single spin on each cycle.  The dynamics within each spin chain is a simple shift, and the  cycle's length, $L$ is  a random variable chosen  from a power-law distribution with power $-\alpha$, i.e. 
\be
P(L) =  \nu/L^\alpha;  ~~~~L_{\min} \leq L \leq L_{\max}
\label{PofL}.
\ee
Here $L_{\min}$ and $L_{\max}$ are the minimal and maximal endpoints of the interval on which the distribution is defined, while $\nu$ is the normalization constant. This model, which we call the Randomly Connected Cycles Network (RCCN), is illustrated in Fig.~\ref{Fig2} 

In this model, the application of external stress is realized by a magnetic field that polarizes the spins and sets them into an ``OFF" position that represents an inactive state of the corresponding cell's component. Soon after the magnetic field is turned off, the magnetization
of the system decays back to its normal state. The first time it crosses zero is assumed to reflect the time point when the cell returns to its normal state. It is associated with the first division time of the cell after starvation, which is the quantity that is measured experimentally. Thus, the primary task of this work is to calculate the first passage distribution of the magnetization and compare it to the first division time distribution of the cells after the starvation period. 

The intention of this paper is as follows: First, we write the equations that describe the dynamics of the RCCN model. Then we present the results of its numerical simulations that reproduced, qualitatively,  the experimental data shown in Fig.~\ref{Fig1}. Next, we analyze the system within a mean-field framework and use the results of this analysis to develop a simple phenomenological description of the problem. The latter will allow us to characterize the distribution function of the total magnetization of the system and obtain an approximate analytical solution for the first passage problem.
 
\section{Definitions and basic equations}
Consider a collection of $N$ closed spin chains that form cycles denoted by the index $i$. Each cycle contains a random number of spins, $L_i$, which we call the cycle length, and each spin may assume one of two values, $\pm 1$ . We denote by  $s_i^{(k)}$ the $k$-th spin on the $i$-th cycle and choose  $s_i^{(0)}$ to be the spins that form couplings among  all cycles (one spin in each cycle). 

Within each cycle the time evolution is a simple shift dynamics,
\begin{subequations}
\label{GE}
\be
s_{i}^{(k)}(t+1)= s_{i}^{(k-1)}(t),~~k=1,2,\cdots,L_i-1 
\ee
while the connecting spins satisfy the equation
 \be
s_i^{(0)}(t+1)= \!\mbox{sign}\left[s_i^{(L_{i}\!-\!1)}(t)\!+\!\sum_{j\neq i} J_{ij} s_j^{(0)}(t)\!+\! h(t) \right].
\ee
\end{subequations}
Here $h(t)$ is the external magnetic field, and $J_{ij}$ are random coupling constants with zero mean and a constant variance inversely proportional to the number of cycles,
thus
\be
\langle J_{ij} \rangle =0,~~\mbox{and}~~~  \langle J_{ij}^2 \rangle = \frac{\gamma^2}{N} ,\label{Jij}
\ee
where $\gamma$ is a constant that characterizes the coupling strength. 
The magnetization of a cycle is defined to be:
\be
M_i(t)=\frac{1}{L_i}\sum_{k=0}^{L_i-1} s_i^{(k)}(t), \label{CycleMag}
\ee
and the total magnetization of the system is
\be
M(t)= \frac{1}{N} \sum_{i=1}^N M_i(t). \label{TotM}
\ee
In a large system with sufficiently strong coupling among the cycles, ($\gamma \gtrsim 1.3$),  for a typical realization of $J_{ij}$, and random initial conditions of the spins, the magnetization,  $M(t),$ behaves like a random process. The magnetic field application increases its average value, which decays back to zero after the magnetic field is turned off. 

The main quantity of interest is the distribution of times, $t$, of first crossing, $M(t)=0$, after magnetic field application. This distribution, denoted by $\rho(t)$, is assumed to reflect the distribution of the first cell division time after the starvation period. In order to avoid data binning, it is convenient to present the results using  the survival probability,
\be
S(t)= \int_t^\infty dt'\rho(t'),\label{survival_prob}
\ee 
 also denoted by 1$-$cdf.

\begin{figure}
\hspace{0.0cm}\includegraphics[width=0.9\columnwidth]{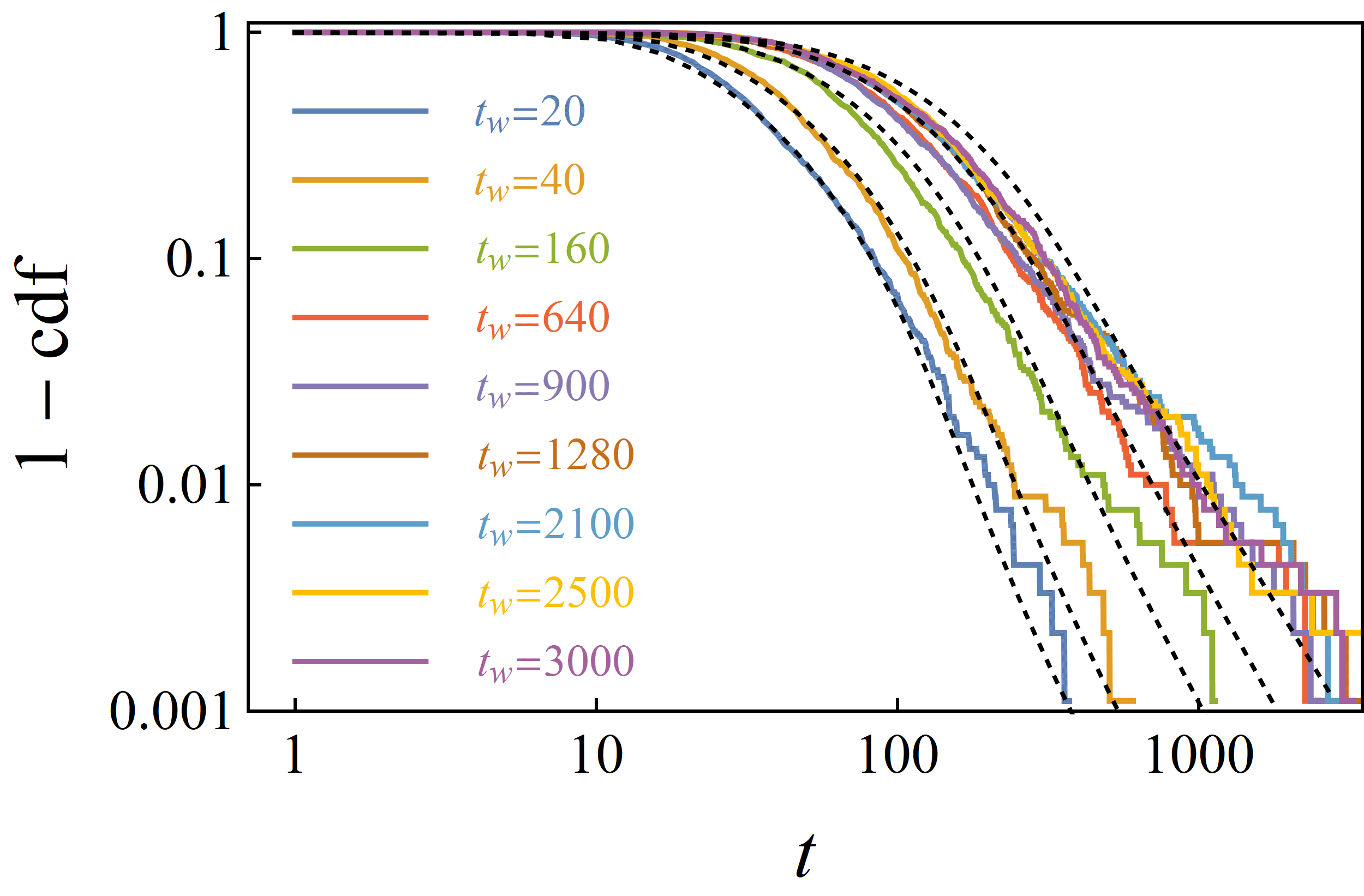} 
\caption{
The magnetization survival probability obtained from simulations of the RCCN model. The black dashed lines are the approximate solution given by formula~(\ref{main}). The dependence of these graphs on the starvation period $t_w$ reflects aging behavior, which saturates for large values of $t_w$. } 
\label{Fig3}
\end{figure}

\section{Numerical simulations.} 
The simulation procedure employed for the study of the RCCN  runs along with the following steps: First, we choose the lengths of the cycles from the distribution (\ref{PofL}). Then we set the coupling constants $J_{ij}$ to be independent random variables from a normal distribution with mean and variance given by Eq.~(\ref{Jij}). Next, Eqs.~(\ref{GE}) are iterated for several (2000) time steps in order to relax the system to some typical state. A constant magnetic field is applied for a period  $t_w$, after which we continue to iterate Eqs.~(\ref{GE}) and record the first time $M(t)$ crosses zero. The statistics are gathered from different simulations in a similar amount to the number of cells monitored in the experiment, and the first crossing time data is used to construct the survival probability. 

The choice of different $J_{ij}$ at each run implies that the survival probability results from ensemble averaging rather than averaging over initial conditions. Because the system is finite (about 400 cycles), by this way, we avoid the problem of choosing some non-typical realization of the coupling constants.

In Fig.~\ref{Fig3}, we present the results of numerical simulations for the survival probability for various stress periods, $t_w$. These are obtained for a system with $2^{14}$ spins, $\gamma=3/2$, $\alpha=3/2$, $L_{\max}=2500$, $L_{\min}=1$, and $h=0.8$ during the stress period. 
These results are in qualitative agreement with the experimental data shown in Fig.~\ref{Fig1}  as they exhibit the following three features: (a) An approximate power-law decay within a wide time interval. (b) The delay in the relaxation increases with the stress period, $t_w$. (c) The behavior of the system exhibits saturation at long stress periods.

\begin{figure}
\hspace{0.0cm}\includegraphics[width=0.9\columnwidth]{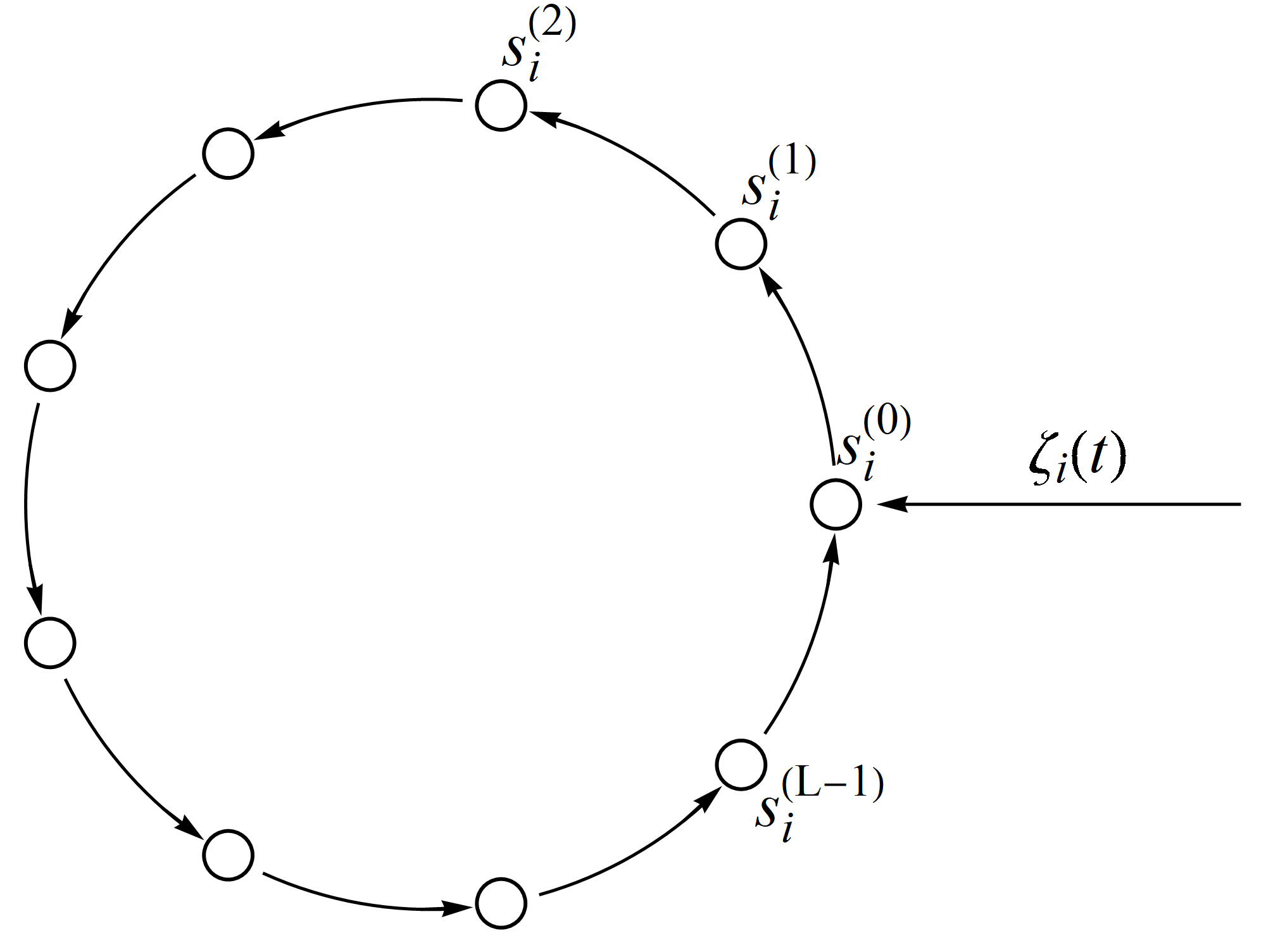} 
\caption{A single cycle of the system. In the mean field approach,
the input from all other cycles is assumed to be noise
whose autocorrelation function is determined self consistently.
 } 
\label{Fig3b}
\end{figure}

\section{Mean Field Approach}

The mean-field approach to the RCCN model is based on the assumption that in a large enough system where all cycles are interconnected, the input signal to a given cycle is a noise term whose statistical properties can be determined self consistently.   

Consider  the $i$-th cycle of the system as illustrated in Fig.~\ref{Fig3b}. Here  the input signal from all the other cycles is
\be
\zeta_i(t)=\sum_{j\neq i} J_{ij}s_j^{(0)}(t). \label{defzeta} 
\ee
When the system is large enough, correlations between cycles are small (see Appendix A) and the input signal is approximately a Gaussian noise with zero mean. Correlations can be determined self consistently from the spins autocorrelation function:
\be 
\langle \zeta_i(t) \zeta_i(t')\rangle =\gamma^2\left\langle 
s^{(0)}_k (t)s_k^{(0)}(t')\right\rangle,\label{noise-spin} 
\ee
where averaging is over the ensemble $J_{ij}$. In a large enough system, ensemble averaging is the same as averaging over the  system's initial conditions. This ergodicity property is discussed in Appendix A. 

From here on, to shorten the notations we shall suppress the cycle index, and denote time using a subscript. With these notations  Eqs.~(\ref{GE}) for the connecting spin in a cycle of length $L$ reduce to 
\be
 s_{t+1}= \mbox{sign}( s_{t-L+1}+ \zeta_t+h_{t}). \label{Eq.Basic}
\ee

It is instructive  to first consider the case where all cycles are of the same length, $L=1$, and zero magnetic field.  Let $p(t)$ be the probability that $s_t=1,$ and assume that the initial condition, $p(0)$, is known. The random noise drives the system into the fixed point $p(t)\to 1/2$. Yet, there will be no change in the spin state until time $t$ if the noise is within the range $-1<\zeta_\nu<1$ for all time steps $\nu=1,2,\cdots t-1$. Therefore
the probability  $p(t)$ evolves according to the  following equation:
\be
p(t)-\frac{1}{2}= V_t \left[ p(0)-\frac{1}{2}\right] \label{pEvolution},
\ee
where, for $t\geq1$,
\be
V_t=\idotsint \limits_{-1< \zeta_\nu<1}~~ \prod_{\nu=0}^{t-1} d\zeta_\nu~  f_t(\zeta_0,\zeta_1 \cdots \zeta_{t-1}) \label{Vt}
\ee
is the probability for no spin-flip for any initial state of the spin within the time interval $(0,t)$. 
Here $f_t(\zeta_0,\zeta_1 \cdots \zeta_{t-1})$ is the joint probability distribution function of the noise at $t$ consecutive time steps.
A formal derivation of this result can be found in Appendix B. 

Let us assume that the input  noise features an exponential decay of correlation:
\be \langle \zeta_t \zeta_{t'} \rangle = \gamma^2 \epsilon^{|t-t'|}, \label{CoMat}
\ee
where $|\epsilon|<1$  is, at the moment, an unknown parameter that determines the time scale of the decay of correlations.

Inverting the covariance matrix of the noise (\ref{CoMat}) results in a tridiagonal matrix that allows one to write the  joint distribution function of the noise in the form
\begin{subequations}
\be 
f_t(\zeta_0,\zeta_1 \cdots \zeta_{t-1})= Z_t \exp \left( -\frac{1}{2} \Phi_t \right),
\ee
where $Z_t$ is the normalization factor, and
\be
\Phi_t =\frac{1}{\gamma^2(1-\epsilon^2)}\sum_{\nu=0}^{t-2} \left(\zeta_\nu-\epsilon \zeta_{\nu+1}\right)^2+\frac{ \zeta_{t-1}^2}{\gamma^2}.  
\ee   
\end{subequations}
The integral (\ref{Vt}) can be evaluated in the limit $\gamma \gg 1$, $|\epsilon| \ll1$ and $1\leq t \ll \gamma^2/ \epsilon^2$ giving
\be
V_t \simeq \mbox{erf}\left(\frac{1}{\sqrt{2}\gamma}\right) \mbox{erf}^{t-1}\left(\frac{1}{\sqrt{2}\gamma\sqrt{1-\epsilon^2}}\right),
\ee
where erf$(x)$ is the error function. 

To identify the self-consistent equation for $\epsilon,$  notice
that $p(t)$ can be interpreted as the conditional probability of $s_t$ given $s_0$. Thus, from $\langle s_t \rangle = 2p(t)-1$ it follows that $\langle s_t s_{0}\rangle=V_{t}$. Substituting this correlation in Eq. (\ref{noise-spin}) and  using Eq.(\ref{CoMat}) we obtain the self consistent equation for $\epsilon$:
\be
\epsilon\simeq -\frac{1}{t} \log V_t = - \log \left[\mbox{erf}\left(\frac{1}{\sqrt{2}\gamma\sqrt{1-\epsilon^2}}\right)\right]
\label{SelfConsistentEpsilon}
\ee
which for large $\gamma$ gives $\epsilon \simeq (2/\pi)^{1/2}/\gamma$. Thus the decay time  of correlations is 
\be
\tau_1=-1/\log \epsilon. \label{tau1}
\ee

In the case where all cycles  are of  length $L>1,$ the dynamics are obtained by rescaling of time, and the relaxation time in this case is 
\be
\tau_L= L \tau_1.
\ee

\begin{figure}[btp]
\vspace{0cm}
\hspace{0.0cm}\includegraphics[width=0.9\columnwidth]{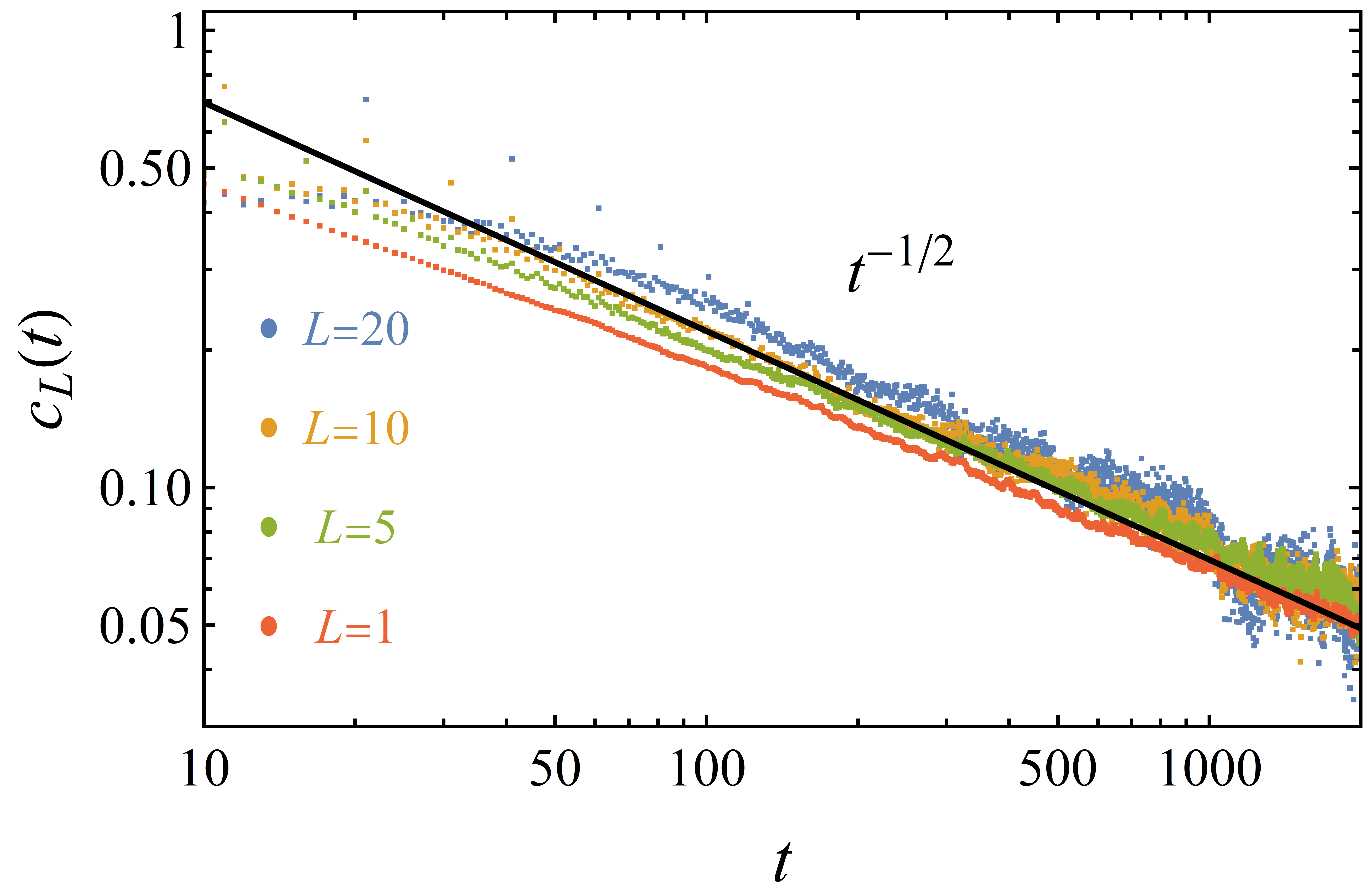}
\caption{The  long time asymptotic  behavior of the  spin correlation function $c_L(t)$ for several cycles lengths, showing a power law decay.  The numerical results  are obtained from 11000 realization of  a system with $2^{14}$ spins, $L_{\max}=2500$, $L_{\min}=1$, and $\alpha=\gamma=3/2$.}
\label{CLfig}
\end{figure}

However, the situation becomes more complicated when the lengths of the cycles are distributed according to the power-law given by Eq. (\ref{PofL}). To reveal the behavior of the noise, in this case, one can try to solve the problem by iteration, starting from an initial approximation where the autocorrelation function of a  spin in a cycle of length $L$ is $\exp(-t/\tau_L)$. Substituting this approximation in the right hand side of Eq. (\ref{noise-spin}) and taking the average over the cycles length distribution yields a power-law decay of the noise correlations $\langle \zeta_0 \zeta_{t} \rangle \sim t^{1-\alpha}$ within the range $L_{\min}\ll t/\tau_1 \ll L_{\max}$. This correlation can be used to construct the first approximation to the noise distribution function, $f_t(\zeta_0,\zeta_1 \cdots \zeta_{t-1})$. With this approximation, the integral (\ref{Vt}) can be calculated in order to obtain the next approximation for the spins correlations, $\langle s_t s_{0}\rangle=V_{t}$ . Then by iterating this procedure, one can improve these approximations. Notice, however, that once the noise correlation features a slow power-law decay,  the spin correlations exhibit a similar behavior because the spin dynamics are dictated by the noise. Thus the spin correlations in a cycle of length $L$,  
\be
c_{L}(t)=\left\langle s^{(0)}(0) s^{(0)}(t) \right\rangle_L  \label{cldef}
\ee
(where averaging is only over cycles of length $L)$  decays as a power law $ t^{1-\alpha}$ within the range $L_{}\ll t/\tau_1 \ll L_{\max}$.  In Fig.~\ref{CLfig}
we demonstrate  that this is indeed the behavior of $c_{L}(t)$ for  $\alpha=3/2$ where correlations decay as $1/\sqrt{t}$. A full analysis of the behavior of  $c_{L}(t)$ in the limits $t \gg L$ and $|t|< L$ can be found in Appendices B \& C.  

Consider now the case where the system is subjected to a constant magnetic field, $h$. From Eq. (\ref{Eq.Basic}) it follows that this amounts to a shift of the noise average $\langle \zeta_t\rangle$ from zero to a finite value $h$, and repeating the calculation that led to (\ref{pEvolution}) now gives\be
p(t)- \frac{1+m}{2}=V_t(h) \left[p(0)- \frac{1+m}{2}\right].
\ee
Here $V_t(h)$ is given by an integral of the form (\ref{Vt}) but with integration range shifts such that, $ -1<\zeta_\nu+h<1$, and $m$ is the magnetization given by 
\be
m=\frac{\mbox{erfc}\mu_--\mbox{erfc}\mu_+}{\mbox{erfc}\mu_-+\mbox{erfc}\mu_+}
\ee
where erfc$(x)$ is the complementary error function, and 
\be
\mu_\pm= \frac{1\pm h}{\gamma \sqrt{2}\sqrt{1-\epsilon^2}}.
\ee
In the strong coupling limit, $\gamma \gg 1$, and a weak magnetic field, $h< \gamma$, one obtains that the magnetization saturates to
\be
M(t) \xrightarrow[t\to \infty]{} m \simeq \sqrt{\frac{2}{\pi}}\frac{h}{\gamma}.\label{saturation_m}
\ee

Notice that the shift dynamics within each cycle imply that the time scale for polarizing the spins within a cycle of length $L$ is $\tau_L=\tau_1 L$, because the magnetic field affects only the connecting spins to other cycles.

A finite average magnetization changes noise properties. In particular, it reduces the phase space for spin fluctuations. Therefore a transient noise contribution, proportional to the square of the magnetization, is generated at the turn "ON" or turn "OFF" of the external magnetic field. This contribution can be neglected when the external magnetic field is small or when the considered time interval is sufficiently far after the point where the magnetic field has been turned off.

In principle, the above description applies only in the limit of large systems $N \gg 1$,  and strong coupling $\gamma \gg 1$, where relaxation is rapid. To obtain slow relaxation, i.e., large $\tau_1$,  we will consider situations where $\gamma$ is of order one. However, $\gamma$ should be large enough to avoid a situation where the cycles become, effectively, decoupled from each other so that their magnetization freezes out. This limit is where the system behavior is analogous to the glass phase of spin glasses. To avoid this phase, $\gamma$ should be above 1.3. This value is obtained from numerical simulations of the system. For cycles of equal length, $L=1$, a similar value is obtained by requiring that the self-consistent equation (\ref{SelfConsistentEpsilon}) has a solution.  

\section{A Phenomenological Model} 

The mean field approach presented above suggests that the magnetization $M_L(t)$ of each cycle (\ref{CycleMag}) is an independent quantity which behaves similarly to a noisy capacitor. Namely, the magnetization of a cycle of length $L$ satisfies a Langevin equation (i.e. the Ornstein Uhlenbeck process):
\be
\frac{\partial M_L(t)}{\partial t}=-\frac{ M_L(t)- \tilde{h}(t)}{\tau_{L}}+ \xi_{L}(t), \label{OUP} 
\ee
where $\tilde{h}$ and $\xi_{L}(t)$ are the effective magnetic field and the applied noise, respectively.  In the strong coupling limit, $\gamma \gg 1$, and weak magnetic field, $|h| \ll 1$,  the relation between $\tilde{h}$ and the magnetic field $h$ is given by equation (\ref{saturation_m}):   $\tilde{h}(t)=\sqrt{2/\pi}h(t)/\gamma$. The  noise, $\xi_{L}(t)$, is  approximated by a random Gaussian noise with zero mean and short range correlations
\be
  \langle \xi_{L}(t) \xi_{L}(t') \rangle= \frac{2\sigma_{L}^2}{\tau_L} \delta(t-t') \label{sigmaL}
  \ee
 where the constant  $\sigma_{L}^{2}$ is determined such that the variance of the magnetization obtained from the mean field approach equals that obtained from the solution of the Langevin Eq. (\ref{OUP}). The calculation of $\sigma_L$ can be found in Appendix D, see Eq.~(\ref{sigmaL2}).

Taking the cycles to be independent, the distribution, $P(M,t)= \langle \delta [M-M(t)] \rangle$, of the total magnetization (\ref{TotM}), is normal  by the central limit theorem:
\be
P(M,t)= \frac{1}{\sqrt{2\pi}\sigma(t)}\exp \left[- \frac{[M-\overline{M}(t)]^2}{2\sigma^{2}(t)}\right] .\label{PofMt}
\ee
The mean magnetization at time $t$, $\overline{M}(t)$, and the variance, $\sigma^2(t)$, can be calculated  using Eq. (\ref{OUP}).

 Assuming the external magnetic field to be constant, $\tilde{h}$, within the time range $(-t_w,0)$ and zero otherwise, the solution of Eq. (\ref{OUP}) for the magnetization of a cycle of length $L$ is
\be
M_L(t)= \overline{M}_L(t)+ \delta M_L(t),
\ee
where for $t>0$,
\be
\overline{M}_L(t)= \tilde{h}\left[1- \exp\left(-\frac{t_w}{\tau_L}\right)\right]\exp\left(-\frac{t}{\tau_L} \right) \label{CyclesMag}
\ee
is the average magnetization, while the fluctuating component is 
\be
\delta M_L(t)= \int_{-t_*}^tdt'\xi_{L}(t')\exp\left(-\frac{t-t'}{\tau_L}\right). \label{Magfluc}
\ee
Here $-t_*$ is the initial time where the dynamics started.

The total average magnetization of the system is now obtained by averaging Eq.~(\ref{CyclesMag}) over the distribution of  cycles length (\ref{PofL}):
\be
\overline{M}(t)=\left\langle \overline{M}_L \right\rangle_L.  
\ee 
This average can be expressed in terms of the exponential integral function, $E_n(x),$
\begin{align}
 \overline{M}(t) \!&= \!\nu \tilde{h}\left\{\frac{1 }{L_{\max}^{\alpha-1}}\!\left[E_{2-\alpha} \!\left(\frac{t}{\tau_{\max}}\right)\!-\!E_{2-\alpha} \left(\frac{t\!+\!t_w}{\tau_{\max}}\right)\right],
\right.\nonumber\\
&-\left.\frac{1 }{L_{\min}^{\alpha-1}}\!\left[E_{2-\alpha} \!\left(\frac{t}{\tau_{\min}}\right)\!-\!E_{2-\alpha} \left(\frac{t\!+\!t_w}{\tau_{\min}}\right)\right]\right\}
 \label{AVMAG}
\end{align}
where $\nu$ is the normalization of the cycles length distribution function (\ref{PofL}), $\tau_{\max}= \tau_1 L_{\max}$, and $\tau_{\min}= \tau_1 L_{\min}$.  

\begin{figure}[btp]
\vspace{0.2cm}
\hspace{0.0cm}\includegraphics[width=0.9\columnwidth]{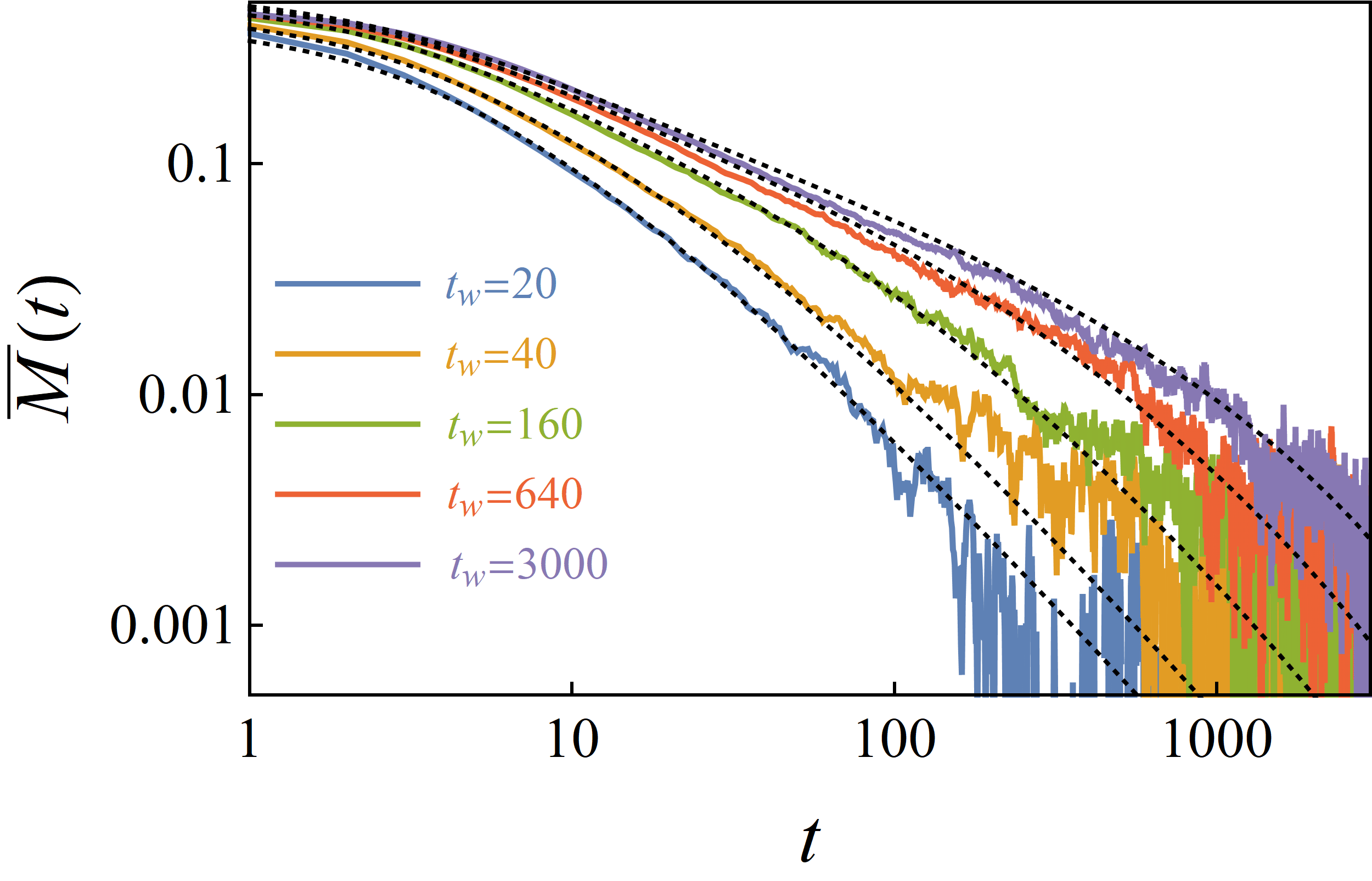}
\caption{A Log-log plot of the average magnetization of the RCCN model for various waiting times. The colored lines are the results of simulation with parameters as in Fig.~\ref{Fig3}, while the dashed lines are the curves obtained from formula (\ref{AVMAG}). (Only 5 values of waiting times, $t_w=20,40,160,640$, and 3000, are presented for clarity).} 
\label{Fig_avg_mag}
\end{figure}

The curves that describe the average magnetization, $\overline{M}(t)$, in a system with $2^{14}$ spins, $\alpha=\gamma=3/2$, $L_{\min}=1$, $L_{\max}=2500$, and $h=0.8$, are depicted in Fig.~\ref{Fig_avg_mag} for various values of waiting times $t_w$. The colored lines show the results of numerical simulations, while the dashed lines are the analytic results of  Eq.~(\ref{AVMAG}) obtained when substituting  $\tilde{h}=0.57$ and $\tau_1=1.9$. These values are slightly different from those obtained from the large $\gamma$ asymptotic expressions  which are  $\tau_1\simeq 1.68$ and $\tilde{h}\simeq 0.43$, because $\gamma$ is not large and the external  magnetic field, $h$, is not weak.
 
The variance of the fluctuations of the magnetization is obtained by squaring Eq.~(\ref{Magfluc}) and averaging both over the noise $\zeta_L(t)$ and over the cycles length distribution, while taking into account that we average over $N$ cycles. By neglecting the effect of magnetic field on the noise variance and setting $t_* =0$, we obtain:
\be
\sigma^2(t)=\frac{1}{N}\left\langle \sigma_{L}^2\left[1-\exp\left( \frac{-2 t}{\tau_L}\right)\right]\right\rangle_L.
\ee 
For our purpose in the next section we shall need only the saturated value of this variance (at long time):
\be
\overline{\sigma}^{2}=\frac{\langle \sigma^{2}_L\rangle}{N}.\label{SigmaInfinity}
\ee 
(see Eq.~(\ref{sigmabar2}) in the Appendix. D).

\section{The survival probability}
\begin{figure}[btp]
\vspace{0.2cm}
\hspace{0.0cm}\includegraphics[width=0.9\columnwidth]{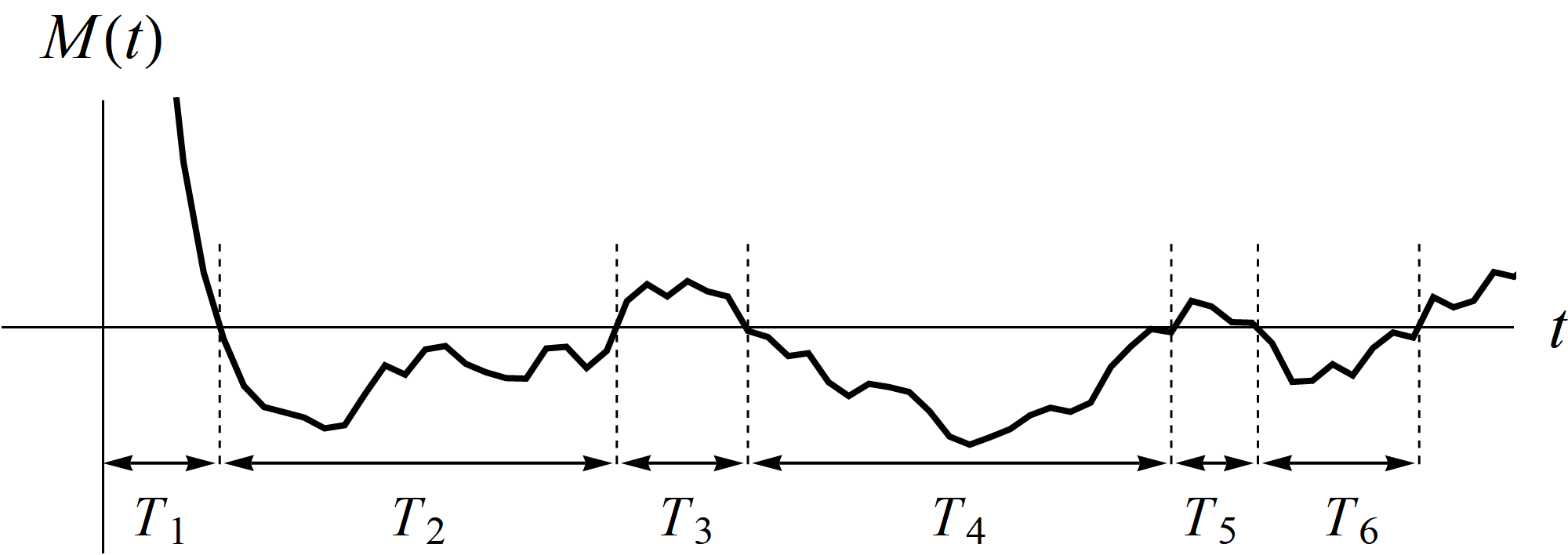}
\caption{An illustration of the typical fluctuating behavior of the magnetization (near the zero baseline) as it decays from a positive initial value.} 
\label{Fig6}
\end{figure}

The typical behavior of a single realization of the magnetization after turning the magnetic field off is illustrated in Fig.~\ref{Fig6}. It exhibits strong fluctuations on top of slow decay. Due to these fluctuations, the magnetization will cross the zero line base downwards and upwards in an alternating manner. Our goal is to calculate the survival probability, $S(t)$ (or 1$-$cdf), i.e., the probability that the magnetization did not cross the zero base line up to time $t$. Its relation to the distribution of times that the magnetization first passes  the zero line value, $\rho(t)$, (i.e. the distribution of   $T_1$ in Fig.~\ref{Fig6}) is described in Eq. (\ref{survival_prob}) and can be also written as
\be
 \rho(t)= - \frac{\partial S(t)}{\partial t}.  \label{rhoS}
\ee

This problem is different from the crossing problem in the Ornstein-Uhlenbeck process\cite{Uhlenbeck1930} because it involves strong temporal correlation of the magnetization . 
Following \cite{Nyberg16} we define $P_<(t)$ to be the probability that $M(t)<0$ at time $t$. From Eq.~(\ref{PofMt}) it follows that
\be
P_<(t)=\int_{-\infty}^{0} dMP(M,t) =\frac{1}{2}
\mbox{erfc}\left[\frac{\overline{M}(t)}{\sqrt{2}\sigma(t)}\right].
\ee

To calculate the survival probability, we also need the distribution functions, $\psi_\pm(\Delta t)$, of the time, $\Delta t,$ that the magnetization returns to zero after it crossed from above and stayed negative (``$-$") or crossed from below and stayed positive (``$+$"). In terms of Fig.~\ref{Fig6}, $\psi_-(\Delta t)$ is the distribution of the even time intervals $T_2$, $T_4$, $T_6$ etc., while ,$\psi_+(\Delta t)$ is the distribution of the odd time intervals $T_3$, $T_5$, $T_7$ etc. (Notice that the distribution function for the first crossing at $T_1$, is special). The main assumption that we need here is that all these time intervals are independent random variables described by a distribution that changes adiabatically in time.  This assumption is based on our numerical simulations, which show that the average magnetization changes very slowly compared to the typical time intervals, $T_k$. 

For a system with discrete-time evolution, the precise value of the crossing time may be defined by interpolation, and the distribution functions $\psi_\pm(\Delta t)$ should be averaged over the distribution of the trajectory overshoot near the crossing point. Here we avoid this complexity by setting our time resolution to be larger than a single time step (the typical return time is of order 50-time steps); hence the magnetization of a single realization is regarded as a continuous function.

Now, let us define $p_{2k-1}(t)$ to be the probability that a trajectory of $M(t)$ starting at $M(0)>0$ ends below the zero baseline at time $t$ after $2k-1$ crossing events (where $k\geq1$). Then:
\be
P_<(t)=\sum_{k=1}^\infty p_{2k-1}(t)\label{P_less}
\ee

It would be instructive to examine $p_{2k-1}(t)$. If  the first crossing event took place at time $t_1<t$ and  there were no other crossing events until time $t$, then
\be
p_1(t)=\int_0^t dt_1 \rho(t_1)Q(t-t_1),
\ee
where $Q(t)$ is the probability that $M(t')$ remains negative for time $t$. This is the probability of no crossing  until time $t$, therefore it can be written in terms of the return probability, i.e.
\be 
Q(t)= 1- \int_0^t dt' \psi_-(t').
\ee  

Consider now the next function $p_3(t)$. Assuming the downward crossing took place at $t_1$, the upward crossing at $t_2$ and the next downward crossing at $t_3$ so that $0\leq t_1 \leq t_2 \leq t_3 \leq t$ we have,
\begin{align}
p_3(t)=\int_0^t dt_1 \rho(t_1)\int_{t_1}^t dt_2 \psi_-(t_2-t_1)\\
\times\int_{t_2}^t dt_3 \psi_+(t_3-t_2)Q(t-t_3).\nonumber
\end{align}
Similar convolution integrals describe $p_{2k-1}(t)$ with $k>2$. Taking the Laplace transform of these formulas we have
\be
\hat{p}_{2k-1}(s)= \hat{\rho}(s) [\hat{\psi}_-(s) \hat{\psi}_+(s)]^{k-1}\hat{Q}(s), \label{hatpk}
\ee
were $\hat{f}(s)$ denotes the Laplace transform of $f(t)$.  

From the above definitions it follows that 
\be
P_<(t)=\sum_{k=1}^\infty p_{2k-1}(t),
\ee
and taking the Laplace transform of equation (\ref{P_less}), using (\ref{hatpk}), and solving for $\hat{\rho}(s),$ we obtain:
\be 
\hat{\rho}(s)=s \hat{P}_<(s)\frac{1-\hat{\psi}_-(s)\hat{\psi}_+(s)}{1-\hat{\psi}_-(s)}
.\ee
Finally,  from the above equation and the  Laplace transform of Eq. (\ref{rhoS}), with $S(0)=1,$ we obtain
\be
\hat{S}(s) =\frac{1}{s}-\hat{P}_<(s)\frac{1-\hat{\psi}_-(s)\hat{\psi}_+(s)}{1-\hat{\psi}_-(s)}.
\label{hatSBasic}
\ee
In order to proceed, one needs the functional form of $\psi_\pm(t)$. These functions are not known and  difficult to compute. Here we assume that each of these functions is characterized by a single time-scale and has an exponential form:
\be
\psi_\pm(t)= \frac{1}{\tau_\pm}\exp\left(-\frac{t}{\tau_\pm}\right),
\ee  
where $\tau_\pm$ are the decay time scales. Taking the Laplace transform of these function and substituting them in Eq.~(\ref{hatSBasic}) we obtain:
\be
\hat{S}(s) =\frac{1}{s}-\hat{P}_<(s)-\frac{\tau_+}{\tau_-}\frac{1}{1+\tau_+ s}\hat{P}_<(s).
\label{hatSexp}
\ee

In order to identify the ratio $\tau_+/\tau_-,$ we multiply  the above equation by $s$ and take the limit $s \to 0$.
 Then using the following property of Laplace transform: $\lim_{s \to 0} s \hat{f}(s) = \lim_{t\to \infty} f(t)$, and the assumption of adiabatic evolution, namely that the magnetization $\overline{M}(t)$ and the standard deviation $\sigma(t)$ can be assumed to be constants over large time intervals (compared to $\tau_\pm$), we obtain:
\be
\frac{\tau_+}{\tau_-}=\frac{1}{P_<(t)}-1
\ee

Thus, the inverse Laplace transform of Eq.~(\ref{hatSexp}), leads to
\be
S(t)= 1-P_<(t) -\left( \frac{1}{P_<(t)}-1\right) \int_0^t  \frac{dt'}{\tau_+} e^{ -t'/\tau_+}P_<(t-t'). \label{main}
\ee

The decay time $\tau_+$  is unknown. From the adiabatic assumption it is clear that it should reduce slowly in time because as the average magnetization becomes smaller the probability for $M(t)$ to stay positive reduces.  Moreover, ageing suggests that  $\tau_+$   also has a weak dependence on the waiting time, $t_w$ which sets the value of the average magnetization, see Appendix E. The dashed lines in Fig.~\ref{Fig3} (corresponding to $t_w=20,40,160,640$ and 3000) are obtained from formula (\ref{main}) with logarithmic dependence on the waiting time:  $\tau_+ = 11\log (t_w/2)$ (see Appendix E).  The other parameters are $\tau_1=1.9$, $\tilde{h}=0.57$ (the same  as for the average magnetization shown in Fig.~\ref{Fig_avg_mag}), and $\sigma(t)\simeq \overline{\sigma}=0.047$ (see Appendix D).

\section{Conclusion} 

To conclude, we constructed a minimal toy model that reproduces the statistical behavior of the division time of a {\it E.\,coli} cell that undergoes a period of acute stress in which its growth is arrested. This construction rests on the following central assumptions: First, the cellular network may be considered random when taking the cell far from its adaptive regime. Second, one can describe this network using Boolean variables that account for each cell component's active or inactive states. Third, the statistical properties of the dynamics of the system are captured by discrete-time evolution. Fourth, the interactions among components can be approximated by two-body interactions. Namely, the state of a spin depends (nonlinearly)  only on sums of the form $\sum_j J_{ij} s_j$ and not, e.g., terms of the form $\sum_{j,k} J_{ijk} s_j s_k$ that represent three-body interactions. With these assumptions, the simplest model we could construct that reproduces the experimental results is the RCCN model illustrated in Fig.~\ref{Fig2}. 

Clearly, none of these assumptions hold on a microscopic level description: The cellular network is not random, as it evolved during billions of years of evolution; the concentration of cell components and the time evolution are continuous; and reactions within the cell usually involve few-body interactions.  Nevertheless, the agreement between the experimental data and the numerical and analytical solution of the model suggests that the RCCN model provides an effective statistical description of the cell. In other words, one expects that a realistic microscopic model of the cellular network can be reduced to the RCCN model in the proper regime of parameters. As such, this model forms a starting point on which a more refined description can be built by taking into account particular features of the cellular network.

Moreover, our study shows that the random statistical behavior of the cells holds valuable information about the cellular network structure. For instance, comparing the solution of the RCCN  model with the experimental data allows one to extract information about the cycles length distribution, e.g., the power $\alpha$ in Eq. (\ref{PofL}), as well as the scale of the longest cellular cycle, $L_{\max}$ of the cell.

In order to see how $\alpha$ is reflected in the survival probability, consider the intermediate asymptotic behavior of $S(t)$ within the range $\tau_{\min} \ll t \ll \tau_{\max}$ which can be evaluated assuming $\tau_+ \ll t$.  Within this range the upper limit in the integral (\ref{main}) can be extended to infinity, while the function $P_<(t-t^\prime)$ can be expanded in powers of $t^\prime$. The zeroth order term of this expansion cancels the first term on the right hand side of Eq. (\ref{main}), while the first order term yields,
\begin{equation}
S(t) \approx  \left\{  
\begin{array}{lr}  b_{+} \left(\frac{\tau_{\max}}{t}\right)^{\alpha+1} & \tau_{\min} \ll t_w \ll t \ll \tau_{\max} \\
  b_{-}\left(\frac{\tau_{\max}}{t}\right)^{\alpha} & \tau_{\min} \ll t \ll \tau_{\max} <t_w \end{array} 
 \right.  , \label{MainAssy}
\end{equation}
with
\be
b_\pm= (\alpha-1) \Gamma(\alpha) \frac{\tilde{h}}{\sqrt{2\pi} \overline{\sigma}}\left(\frac{L_{\min}}{L_{\max}}\right)^{\alpha-1}
\frac{\tau_+}{\tau_{\max}} \left(\frac{\alpha t_w}{\tau_{\max}}\right)^{\mu_\pm}
\ee 
where $\mu_+=1$, and $\mu_-=0$. 

Eq. (\ref{MainAssy}) shows that  the distribution of the cycles length (\ref{PofL}) determines the intermediate asymptotic behavior of the survival probability: The power-law decay changes from $t^{-\alpha-1}$ for short waiting time to $ t^{-\alpha}$ at saturation, i.e. when $t_w \geq\tau_{\max}$. Notice that this result is independent of any other parameter of the system.  Thus, the behavior of the survival probability at saturation reflects the distribution of the lengths of the cycles \cite{kaplan2021observation}.

From a theoretical viewpoint, the RCCN model is a complicated model, and the phenomenological description that we have presented here provides only a basic description.  For example, we did not take the effect of magnetization on the noise level, the precise dependence of $\tau_+$ on time, and we also did not account for possible cross-correlations between different cycles whose cumulative effect, despite being very small, may become significant.
 
The theoretical framework provided by the RCCN model opens the possibility to study many other universal features of the system such as correlations between cycles; the effect of stress focused on a single node or a limited number of them; and the sensitivity of the dynamics to a small change in the initial conditions. These questions are left for future study.  

\section{Appendix A: Ergodicity}

Eq.~(\ref{noise-spin}) is straightforwardly obtained by ensemble averaging.  In the first part of this Appendix, we show that the same equation can also be derived by averaging over initial conditions of the spin configurations. However, this derivation relies on the assumption that different cycles are uncorrelated. In the second part of the Appendix, we show that the correlation between cycles decay as $1/\sqrt{N}$ where $N$ is the number of cycles.     

In what follows we use overbar to denote averaging over initial conditions and over the cycles. Using definition (\ref{defzeta}), we have 
\be
\overline{\zeta_i(t)\zeta_i(t')}=\frac{1}{N} \sum_i \sum_{k\neq i}\sum_{k'\neq i} J_{ik} J_{ik'}\overline{s_k(t) s_{k'}(t')}.
\ee
Hereinafter we suppress the upper index that denote the connecting spins in each cycles. Assuming the cycles to be uncorrelated implies that only terms with $k=k'$ contribute to the above sum. This diagonal approximation yields
\be
\overline{\zeta_i(t)\zeta_i(t')}=\frac{1}{N} \sum_i\sum_{k\neq i} J_{ik}^2\overline{s_k(t) s_{k}(t')}   
\ee
and by the law of large numbers, in the limit $N \to \infty$, we obtain 
\be
\overline{\zeta_i(t)\zeta_i(t')}=\gamma^2\overline{s_k(t) s_{k}(t')}.
 \ee

Comparing this formula with Eq.(\ref{noise-spin}) shows that averaging over initial conditions is equivalent to ensemble averaging provided $\overline{s_k^{(0)}(t) s_{k'}^{(0)}(t')} \to 0$ when $N \to \infty$ and $k \neq k'$.  To  validate this property consider  the following quantity: 
\be
\Phi_{ij}=\left\langle J_{ij}\overline{s_i(t) s_{j}(t)}\right\rangle,~i \ne j. \label{Phi_ij}
\ee
Expressing $s_i(t)$ in terms of its previous time step, using Eqs.~(\ref{GE}) we have
\be 
\Phi_{ij}=\left\langle J_{ij}\overline{\mbox{sign}[s_{i}^{}(t-L_i)+
 \zeta_i(t-1)]s_j^{}(t)}\right\rangle, 
 \ee
where $\zeta_i(t-1)$ is the input signal to the $i$-th cycle \label{def:zeta}. Let us separate from this signal the contribution that comes from the $j$-th spin, $\zeta_i(t)= J_{ij} s_j(t) + \tilde{\zeta}_i(t)$, and expand sign function to linear order in $J_{ij}$ (this expansion is justified because $J_{ij}$ is of order $1/\sqrt{N}$ and we consider the limit $N \to \infty$). Thus 
\begin{eqnarray}
\Phi_{ij}&&=\left\langle J_{ij}\overline{\mbox{sign}[s_{i}(t-L_i)+  J_{ij} s_j(t-1) + \tilde{\zeta}_i(t-1)] s_j(t)} \right\rangle\nonumber \\
&&=\left\langle J_{ij}\overline{\mbox{sign}[s_{i}(t-L_i)+ \tilde{\zeta}_i(t-1)]
s_j(t)}\right\rangle \label{CoExpand}\\
&&+\left\langle J_{ij}^2\overline{s_j(t-1)s_{j_{}} (t)2\delta[s_{i}(t-L_i)
+ \tilde{\zeta}_i(t-1)]}\right\rangle  \nonumber 
\end{eqnarray}

Now let us assume, self consistently, that  the spins are uncorrelated. Then $\tilde{\zeta}_i(t)$ is a Gaussian noise,  independent of $s_i$ and $s_j$, with zero mean and with variance $\gamma^2$; hence, taking first the mean over the noise only,
$\overline{\mbox{sign}[s_{i}(t)+ \tilde{\zeta}_i]}=s_{i}(t)\overline{\mbox{sign}[1+ \tilde{\zeta}_i]}$ and the term in the middle line of Eq.~(\ref{CoExpand}) vanishes upon averaging. Similarly $\overline{\delta[s_{i}(t)+ \tilde{\zeta}_i]}= P[-s_{i}(t)]$ where  $P(\tilde{\zeta}_i)$ is the noise distribution function. Taking into account that the maximal value of this distribution function is $1/(\sqrt{2 \pi} \gamma)$ and that the maximal value of $s_{j}(t)$ is $1$, we obtain
\be
\Phi_{ij} \leq\left\langle J_{ij}^2\max [P(\tilde{\zeta})]\right\rangle = \frac{ \gamma}{\sqrt{2 \pi} N}
\ee
Noting that the spatial correlations is fully determined by $J_{ij}$, 
This result together with the definition of (\ref{Phi_ij}) implies that
\be
\overline{s_i(t) s_{j}(t)} \sim \frac{J_{ij}+J_{ji}}{\gamma}\sim \frac{1}{\sqrt{N}}.
\ee

\section{Appendix B: Derivation of Eq.~(\ref{pEvolution})}

Consider the mean field approximation for the RCCN model  where all cycles are of length $L=1$. The spin dynamical equations are:
\be
s_{t+1}= \mbox{sign}(s_t+ \zeta_t)
\ee
where $\zeta_t$ is the input noise.  Let us look first on the case where the noise is uncorrelated.  Then the probability, $p(t)$ that $s_t=1$ satisfies the equation:
\be
p(t+1)= p(t) P( \zeta_t> -1)+ [1-p(t)] P(\zeta_t>1) \label{SimpleEvolution}
\ee
Here the first term describes the probability that $s_t=1$
 multiplied by the probability that the noise cannot flip it, i.e., the probability that $\zeta_t>-1$. The second term describes the case where $s_t=-1$ but the noise is sufficiently large to flip it. Next, we use the following general properties of the noise probabilities:
\begin{subequations}
\label{ProbProp}
\be
P( \zeta_t> -1)= q+ P(\zeta_t>1) 
\ee
where 
\be
q= P( -1<\zeta_t<1),
\ee 
and 
\be
P(\zeta_t>1)+P(\zeta_t<-1)+ q= 2P(\zeta_t>1)+q=1,
\ee
\end{subequations}
where here we use the property that the noise distribution is an even function, i.e. $ P(\zeta_t>1)=P(\zeta_t<-1)$. Substituting these equations in (\ref{SimpleEvolution}) gives
\be
p(t+1)-\frac{1}{2} = q \left[p(t)- \frac{1}{2}\right]
\ee

From here Eq.~(\ref{pEvolution}) directly follows with $V_t= q^t$.  

Consider now the noise be correlated and let us calculate the probability that after two steps $s_2=1$. This probability is given by the following sum:
\begin{widetext}
\be
p(2)= P( \zeta_1>1)+ P(-1< \zeta_1<1, \zeta_0>1)+ P(-1<\zeta_1<1, -1<\zeta_0<1)p(0).
\ee
The first term on the right-hand side of this equation is associated with the situation that the noise at the last step is sufficiently large to ensure that the spin will be one independent of its state. The second term describes the case where the noise in the last step is within the  range where no spin flip occur (irrespective of the initial spin state), but the noise in the previous step is large enough to ensure that the spin is in "one" state. Finally, the last term is the probability that the spin was initially at one state, and the noise in both time steps  doesn't flip the spin. These probabilities refer to disjoint sets of events and therefore add up. 

Now, using relations similar to (\ref{ProbProp}) allows one to rewrite the above equation in the form:
\be
p(2)= \frac{1}{2} \left[1-P( -1<\zeta_1<1)\right]+ \frac{1}{2}[P(-1< \zeta_1<1)-V_{2}] +V_{2} p(0),
\ee
\end{widetext}
where
\be
V_2= P(-1<\zeta_1<1, -1<\zeta_0<1).
\ee  
Thus
\be 
p(2)-\frac{1}{2}= V_2 \left[p(0)-\frac{1}{2}\right].
\ee
This derivation can be generalized to any number of time steps giving Eq.~(\ref{pEvolution}).

\section{Appendix C: Noise in the RCCN model}

In this Appendix, we study the fluctuation of the cycles magnetization.  To this end, we shall characterize the correlation function  (\ref{cldef}) of the connecting spin in a cycle of length $L$, and use it to calculate the magnetization variance in order to identify  $\sigma_L^2$, and its average over the distribution of the lengths of the cycles, $\overline{\sigma}^2$.

The essential quantity needed for our analysis is the conditional probability of a connecting spin and the noise acting on the same spin at some different time, $P(\zeta_{t}|s_0)$, where the subscript denotes the time step. It is easy to calculate this conditional probability for $t=-1$ because the time evolution $s_0=\mbox{sign}(s_{-1} + \zeta_{-1})$ allows one to calculate the conditional probability $P(s_0|\zeta_{-1})$. Then using Bayes' theorem, we have
\be
P(\zeta_{t-1}|s_0)= P(s_0|\zeta_{-1}) \frac{P_0(\zeta_{-1})}{P(s_0)} 
\ee
where $P(s_0)=1/2$ is the unconditional probability of the spin (in the absence of magnetic field), and $P_{0}(\zeta)$ is the probability distribution of the noise at some arbitrary time, which is normal distribution with zero mean and variance $\gamma^2$.  From here we obtain that for $s_0=1$:  
\be
P(\zeta_{-1}|1)=\left\{ \begin{array}{cc} 2P_0(\zeta_{-1}) & \zeta_{-1}>1 \\ P_0 (\zeta_{-1})  & |\zeta_{-1}|<1 \\
0  & \zeta_{-1}<-1 \end{array} \right. \label{condm1}
\ee
The conditional probability is symmetric with respect to a change sign of both $s_0$ and $\zeta_{-1}$ hence $P(\zeta_{-1}|-1)=P(-\zeta_{-1}|1)$.

Now we can calculate the conditional probability at any other time by the integral
\be
P(\zeta_{t}|s_0)=\int d\zeta_{-1} P(\zeta_{t}| \zeta_{-1},s_0) P(\zeta_{-1}|s_0). \label{condgen}
\ee
Here $P(\zeta_{t}| \zeta_{-1},s_0)$ is the conditional probability of the noise at time $t$ given the noise value at time $t=-1,$ and the state of the spin, $s_0$. However, the effect of a single spin on the noise  is negligible because the noise results from a very large number of spins; hence   
\begin{eqnarray}
P(\zeta_{t}| \zeta_{-1},s_0)&=& P(\zeta_{t}| \zeta_{-1}) \nonumber\\
&=& 
\frac{\exp \left( -\frac{(\zeta_{t}-\mu_{t+1} \zeta_{-1})^2}{2\gamma^2 (1-\mu_{t+1}^2)}\right)}
{\gamma \sqrt{2\pi(1-\mu_{t+1}^2)}} \label{condnoise}
\end{eqnarray}
where
$\mu_{t+1}= \langle \zeta_{t} \zeta_{-1}\rangle/\gamma^2$ is the noise correlation function normalized by its variance. Substituting (\ref{condm1}) and (\ref{condnoise}) in (\ref{condgen}) yields:
\be
P(\zeta_{t}|s_0)= \left[1+ s_0 \chi(\zeta_{t})\right] P_{0}(\zeta_{t}) \label{cond1}
\ee
where 
\be
\chi(\zeta)=\frac{1}{2} \sum_{\pm}\pm \mbox{erf}\left( \frac{1\pm \zeta
\mu_{t+1}}{\gamma\sqrt{2-2\mu_{t+1}^2}}\right). 
\ee

We turn now to calculate the autocorrelation function of the connecting spin in a cycle of length $L$ at time $t=L$,  $c_L(L)$, in the limit of zero magnetic field. Setting $t=L-1$ in Eq. (\ref{Eq.Basic}),  multiplying it by $s_0$, and taking the average we obtain
\be
c_L(L)=\langle s_0 s_{L}\rangle=\langle s_0\mbox{sign}( s_{0}+ \zeta_{L-1})\rangle. \label{cll}
\ee
This average can be evaluated using (\ref{cond1})  giving
\be 
c_{L}(L)=  \mbox{erf}\left( \frac{1}{\sqrt{2} \gamma}\right)+2\int_{1}^\infty d\zeta \chi(\zeta) P_0(\zeta).
\ee 
This is a general function of  $\mu_L$ and $\gamma$, but for long cycles one can assume that $\mu_L \ll 1$, and the above integral may  be expanded up to linear order in this parameter, giving
\be
c_L(L)\simeq  \mbox{erf}\left( \frac{1}{\sqrt{2} \gamma}\right)+q\mu_L, \label{cLofL}
\ee
where
\be
q= \frac{2}{\pi}\exp\left(-\frac{1}{\gamma^2}\right).
\ee  

Consider, now  the same correlation function but at time $t=2L$, i.e.,  $c_{L}(2L).$ Setting $t=2L-1$ in Eq. (\ref{Eq.Basic}), multiplying by $s_0$ , and expressing $s_L$ in terms of $s_0$ using Eq.(\ref{Eq.Basic}) once again we obtain:
\be
c_L(2L)=\left\langle s_0 ~\mbox{sign}[~ \mbox{sign}(s_0+ \zeta_{L-1})+ \zeta_{2L-1}]\right\rangle. \label{Cll1}
\ee 
To evaluate this quantity, we need the joint distribution of the noise at two points in time conditioned by the state of the spin $s_0$: 
\be
P(\zeta_t, \zeta_{t'}|s_0)= \int d\zeta_{-1}\ P_0(\zeta_t ,\zeta_{t'}|\zeta_{-1})P(\zeta_{-1}|s_0)
\ee 
To leading order in the noise correlations, $\mu_t$, This integral gives:
\be
\frac{P_{0}(\zeta_t, \zeta_{t'}|1)}{ P_0(\zeta_t) P_0(\zeta_{t'})} \simeq 1+\zeta_t \zeta_{t'} \mu_{t-t'}+\frac{\sqrt{q}}{\gamma}(\zeta_t \mu_t+ \zeta_{t'} \mu_{t'}). \label{cond2}
\ee
The average on the right hand side of (\ref{Cll1}) can be evaluated by
expressing the outer sign function in terms of its  Fourier integral,
\be
\mbox{sign}(x)=2\mbox{Re} \int_0^\infty \frac{d\eta}{i\pi\eta} \exp(i\eta x).
\ee 
Preforming the average yields
\be
c_L(2L)\simeq \mbox{erf}^2\left( \!\frac{1}{\sqrt{2}\gamma}\!\right)+q\left[\mu_L \mbox{erf}\left( \frac{1}{\sqrt{2}\gamma}\right)+\mu_{2L}\right].
\ee
Notice that the contribution from the noise autocorrelation, i.e 
the quadratic term on the right-hand side of Eq. (\ref{cond2}), vanishes. Thus one can interpret the above formula in the following way: The first term is simply the probability that the spin does not flip by the noise after two events where its state may have been changed (at time $L$ and time $2L)$, assuming these events to be independent. The two other contributions represent corrections to this result due to correlations of the noise with the spin. The first is the product of the probabilities that noise does not flip the spin at one time event multiplied by the excess probability for the spin to remain in its state due to correlations with the spin after time $L$. The second contribution comes from correlations of the noise and the spin after time $2L$. 

A similar calculation for  $n$ time steps (of length $L$), in the asymptotic limit $L\gg 1$ gives
\be
c_L(nL)\simeq q \sum_{j=1}^{n}  \mbox{erf}^{n-j}\left( \!\frac{1}{\sqrt{2}\gamma}\right)\mu _{jL}+\mbox{erf}^n\left(\! \frac{1}{\sqrt{2}\gamma}\!\right). \label{condLn}
\ee

Assuming this behavior is characteristic for any discrete time $t$ and not only to the values $t=nL$ one can now solve this equation self consistently by averaging $c_L(t)$ over the power law distribution of the cycles length (\ref{PofL}). This average should be proportional to the noise autocoronation $\mu_t$.  

Consider, first, the average of the last term in the above formula  (which decays exponentially).  In the asymptotic limit $\tau_{\min} \ll t\ll \tau_{\max}$  it yields
\be
\left\langle \mbox{erf}^{\frac{t}{L}}\left(\frac{1}{\sqrt{2}\gamma}\right)\right\rangle_L=
\nu \Gamma(\alpha-1)\left(\frac{\tau_{1}}{t}\right)^{\alpha -1}
\ee 
where in the approximation where correlations are taken only to first order to,  $\tau_1^{-1}\simeq -\log\mbox{[erf}( 1/\sqrt{2} \gamma)]$.  This term yield an algebraic decay,  and substituting the same form of decay, $\mu_{jL}\simeq  b (jL)^{1-\alpha}$,  in Eq. (\ref{condLn}),  the sum can be expressed in terms of
Lerch transcendent  which yields the following asymptotic formula:
\be
c_L(t)\simeq \tilde{b}\left( \frac{1}{t+L}\right)^{\alpha-1}
\!\!+\exp\left(-\frac{t}{\tau_{1}L}\right) \label{cLoft}
\ee
where
\be
\tilde{b}=\frac{bq{}}
{1-\mbox{erf}\left(\frac{1}{\sqrt{2}\gamma}\right)}.
\ee
This result shows that in the large time asymptotic limit, all cycles correlation functions decay as a power law with the same power and amplitude as illustrated in Fig.~\ref{CLfig} and Fig.~\ref{bootstrap}.

\section{Appendix D: The magnetization noise}

In this appendix, we calculate the magnetization noise variance, $\overline{\sigma}^2$, at zero magnetic field. It is obtained by averaging the variance of the magnetization noise, $\sigma_L^2$, of cycles of length $L$ over the cycles length distribution (\ref{PofL}). 

Consider the magnetization of a cycle of length $L$. Taking into account that the dynamics within the cycle is shift dynamics, we have
\be
M_L(t)= \frac{1}{L} \sum_{k=0}^{L-1} s^{(k)}(t)=  \frac{1}{L} \sum_{t'=0}^{L-1} s^{(0)}(t-t').
\ee
Thus when the system reaches a stationary state,  the variance of the magnetization of a cycle of length $L$ is given by:
\begin{align}
\langle M_L^2\rangle &= \frac{1}{L^2} \sum_{t''=0}^{L-1}\sum_{t'=0}^{L-1} \langle s^{(0)}(t-t'')s^{(0)}(t-t')\rangle\nonumber \\&= \frac{1}{L^2} \sum_{t''=0}^{L-1}\sum_{t'=0}^{L-1} c_L(t''-t') \nonumber \\&= \frac{1}{L}+\frac{2}{L^2} \sum_{t=1}^{L-1} (L-t)c_L(t),  \label{MLVar}
\end{align}
where we took into account that $c_L(0)=1$. 

\begin{figure}[btp]
\vspace{0.2cm}
\hspace{0.0cm}\includegraphics[width=0.9\columnwidth]{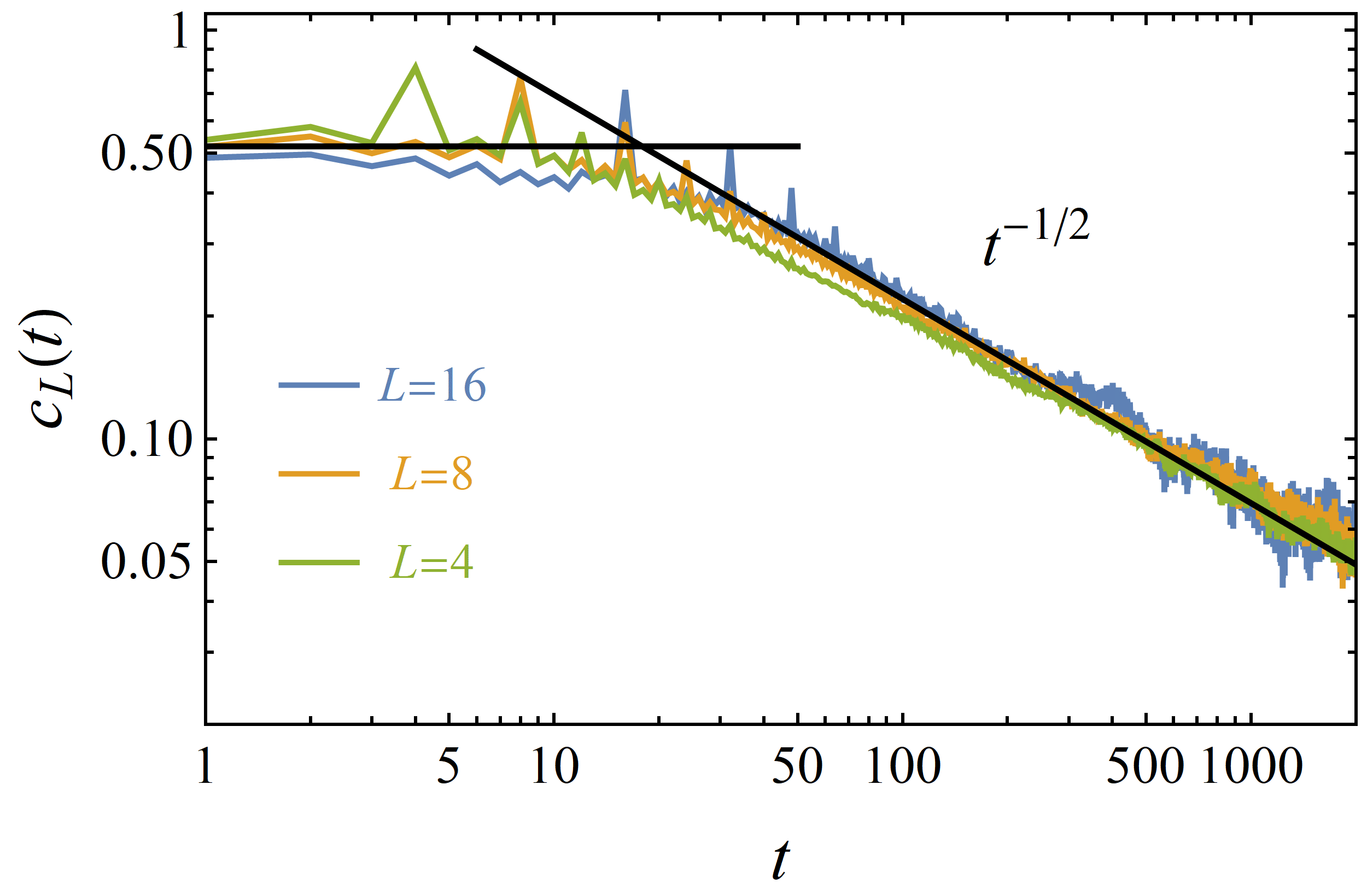}
\caption{A log-log plot of a connecting spin correlation function, $c_L(t)$, for various values of cycles length. The dots are connected by lines for clarity. } 
\label{bootstrap}
\end{figure}

The above sum depends on the behavior of $c_L(t)$ within the short time regime $0<t<L$. The results obtained in Appendix C (see Eqs. (\ref{cLoft})  and (\ref{cLofL})) indicate that within this range, the correlation length is essentially constant.  This conclusion is also supported by the numerical results shown in Fig.~\ref{bootstrap}, where we depict $c_L(t)$ at three values of cycle length for $\alpha=3/2$. This figure shows modulations of the correlations on the scale of $L$  with an approximate constant behavior within the range $1 \leq t\leq L-1$.

If we set $c_L(t)$ to be  constant $k$, then
\be
\sigma_L^2 \simeq\langle M_L^2\rangle  \simeq \frac{1}{L}+ k\frac{L-1}{L} \label{sigmaL2}
\ee

To calculate the  $\overline{\sigma}^2$ in Eq.~(\ref{SigmaInfinity}),
one should take the average of $\sigma_L^2$ over the cycles length distribution. For simplicity we assume $L_{\min}=1$ and extend the sum to infinity (as it converges for $\alpha >1$). With these approximations we obtain
\be
\overline{\sigma}^2=\frac{1}{N}\langle \sigma_L^2 \rangle =\frac{1}{N}\ \left[ k+\frac{(1-k) \zeta(1+\alpha)}{\zeta(\alpha)}\right], \label{sigmabar2}
\ee
where $N$ is the average number of cycles, while $\zeta(x)$ is the Riemann zeta function. 

Finally, it remains to obtain an estimate for the constant $k$. For this purpose, let us calculate the correlation function at one time step,
\be
c_L(1)=\langle s_0 s_1 \rangle= \langle s_0 \mbox{sign} (s_{1-L}+\zeta_1) \rangle.
\ee
In order to calculate this average one needs the conditional probability $P(\zeta_{1}|s_0, s_{1-L})$, namely the distribution of the noise at time $t=1$ for a given state of the spin at previous times  $t=0$ and $t=1-L$. For sufficiently large cycles, one expects that the correlations of the noise with the state of the spin at the latter time can be neglected and to use the approximation $P(\zeta_{1}|s_0, s_{1-L})\simeq P(\zeta_{1}|s_0)$. Then using (\ref{cond1}) to perform the average, and linearizing with respect to $\mu_1$ (although this is not strictly justified because $\mu_1$ is of order one, but here we are only interested in the  approximate value) we obtain:
 
\be
c_L(1)=\left\langle s_0 \mbox{erf} \left( \frac{s_{1-L}}{\sqrt{2} \gamma} \right) \right\rangle + q\mu_1.
\ee
Next, to perform the average over $s_{1-L}$ we use the conditional probability:
\be
P(s_{1-L}|s_0)=\left\{ \begin{array}{ll} \frac{1+u}{2} & \mbox{  for}~~ s_{1-L}=s_0 \\ 
\frac{1-u}{2} & \mbox{  for}~~ s_{1-L}=-s_0
\end{array} \right.
\ee
where $u= \langle s_0 s_{1-L}\rangle = c_L(L-1)$. From the last two equations  we obtain:
\begin{align}
c_L(1) &\simeq c_L(L-1) \mbox{erf} \left( \frac{1}{\sqrt{2} \gamma} \right)+ q\mu_1 \label{boot} 
\end{align}
This formula connects the correlation function at one time step with that at the time step just before the completion of a full cycle  of $L$ steps. Setting
$k=c_L(1)\simeq c_L(L-1)$ and solving for $k$ we obtain.
\be 
k \simeq \frac{q \mu_{1}}{1-\mbox{erf}\left(  \frac{1}{\sqrt{2}\gamma}\right)}.
 \ee
 
In our numerical study $\alpha=\gamma=3/2,$ which implies that $k\simeq 0.81 \mu_1$. Taking  also that average number of cycles is $N=408$, and approximating  $\mu_1\simeq 1$ we obtain from (\ref{sigmabar2}) that $\overline{\sigma}=0.047$ which is the same as the numerical result.
 
\section{Appendix E: The waiting time Dependence of $\tau_+$}

This appendix explains the logarithmic dependence of the return time, $\tau_+,$ on the waiting time, $t_w$.  For this purpose, we take note that formula (\ref{main})  is obtained by taking into account correlations in the relaxation of the magnetizaion and the effect of the second term is manifested only at a sufficiently long time, when the survival probability becomes small. Thus in what follows, we estimate $\tau_+$ by focusing our attention on the regime:
\be
t \sim \tau_{\max}, \quad \mbox{and} \quad \tau_{\min} \ll t_w < \tau_{\max}. \label{regime}
\ee

The magnetization is a sum of two contributions: the mean value of the magnetization and a fluctuating part.  From Eq.~(\ref{AVMAG}) it follows that average magnetization in the regime of interest (\ref{regime}) is given by
\be
\overline{M} \approx c t_w. \label{M}
\ee
where
\be 
c \simeq (\alpha-1) \Gamma(\alpha,1)  \frac{\tilde{h}}{\tau_{\max}} \left( \frac{L_{\min}}{L_{\max}} \right)^{\alpha-1},
 \ee
and $\Gamma(a,z)$ is the incomplete gamma function. This positive contribution comes mainly from the long cycles which decay slowly  and keep memory of the time that the system was subjected to the magnetic field. It is therefore clear that it increases with the waiting time, $t_w$, (as long as it is shorter than the longest cycle).

On the other hand, the main contribution to the fluctuating part of the magnetization comes from short cycles that quickly lose memory of the magnetic field, and because there are many of them. The relaxation of these very short orbits is approximately exponential as follows from (\ref{CyclesMag}).

Now consider a trajectory of the magnetization that crosses zero value upwards.  Since the average magnetization is positive, the fluctuation must be negative with amplitude given by (\ref{M}). Thus one expects it to have the form $\delta M(t) \approx -\bar{M} \exp( -t/\tau_{\min})$, where time is measured from the crossing point. Now, the typical time that the magnetization remains positive is several times (say $n$) of the time that takes this fluctuation to relax. Namely it is obtained from the condition $\delta M( \tau_+/q) =-\delta$ where $\delta$ is some small (positive) value of the magnetization.    Solution of this equation yields a logarithmic dependence on the return time, $\tau_+= n\tau_{\min} \log (c t_w/\delta)$.

\bibliography{main}% Produces the bibliography via BibTeX.

\begin{thebibliography}{45}%
\makeatletter
\providecommand \@ifxundefined [1]{%
 \@ifx{#1\undefined}
}%
\providecommand \@ifnum [1]{%
 \ifnum #1\expandafter \@firstoftwo
 \else \expandafter \@secondoftwo
 \fi
}%
\providecommand \@ifx [1]{%
 \ifx #1\expandafter \@firstoftwo
 \else \expandafter \@secondoftwo
 \fi
}%
\providecommand \natexlab [1]{#1}%
\providecommand \enquote  [1]{``#1''}%
\providecommand \bibnamefont  [1]{#1}%
\providecommand \bibfnamefont [1]{#1}%
\providecommand \citenamefont [1]{#1}%
\providecommand \href@noop [0]{\@secondoftwo}%
\providecommand \href [0]{\begingroup \@sanitize@url \@href}%
\providecommand \@href[1]{\@@startlink{#1}\@@href}%
\providecommand \@@href[1]{\endgroup#1\@@endlink}%
\providecommand \@sanitize@url [0]{\catcode `\\12\catcode `\$12\catcode
  `\&12\catcode `\#12\catcode `\^12\catcode `\_12\catcode `\%12\relax}%
\providecommand \@@startlink[1]{}%
\providecommand \@@endlink[0]{}%
\providecommand \url  [0]{\begingroup\@sanitize@url \@url }%
\providecommand \@url [1]{\endgroup\@href {#1}{\urlprefix }}%
\providecommand \urlprefix  [0]{URL }%
\providecommand \Eprint [0]{\href }%
\providecommand \doibase [0]{https://doi.org/}%
\providecommand \selectlanguage [0]{\@gobble}%
\providecommand \bibinfo  [0]{\@secondoftwo}%
\providecommand \bibfield  [0]{\@secondoftwo}%
\providecommand \translation [1]{[#1]}%
\providecommand \BibitemOpen [0]{}%
\providecommand \bibitemStop [0]{}%
\providecommand \bibitemNoStop [0]{.\EOS\space}%
\providecommand \EOS [0]{\spacefactor3000\relax}%
\providecommand \BibitemShut  [1]{\csname bibitem#1\endcsname}%
\let\auto@bib@innerbib\@empty
%</preamble>
\bibitem [{\citenamefont {Parisi}(1993)}]{parisi1993statistical}%
  \BibitemOpen
  \bibfield  {author} {\bibinfo {author} {\bibfnamefont {G.}~\bibnamefont
  {Parisi}},\ }\bibfield  {title} {\bibinfo {title} {Statistical physics and
  biology},\ }\href@noop {} {\bibfield  {journal} {\bibinfo  {journal} {Physics
  World}\ }\textbf {\bibinfo {volume} {6}},\ \bibinfo {pages} {42} (\bibinfo
  {year} {1993})}\BibitemShut {NoStop}%
\bibitem [{\citenamefont {Ozbudak}\ \emph {et~al.}(2004)\citenamefont
  {Ozbudak}, \citenamefont {Thattai}, \citenamefont {Lim}, \citenamefont
  {Shraiman},\ and\ \citenamefont
  {Van~Oudenaarden}}]{ozbudak2004multistability}%
  \BibitemOpen
  \bibfield  {author} {\bibinfo {author} {\bibfnamefont {E.~M.}\ \bibnamefont
  {Ozbudak}}, \bibinfo {author} {\bibfnamefont {M.}~\bibnamefont {Thattai}},
  \bibinfo {author} {\bibfnamefont {H.~N.}\ \bibnamefont {Lim}}, \bibinfo
  {author} {\bibfnamefont {B.~I.}\ \bibnamefont {Shraiman}},\ and\ \bibinfo
  {author} {\bibfnamefont {A.}~\bibnamefont {Van~Oudenaarden}},\ }\bibfield
  {title} {\bibinfo {title} {Multistability in the lactose utilization network
  of escherichia coli},\ }\href@noop {} {\bibfield  {journal} {\bibinfo
  {journal} {Nature}\ }\textbf {\bibinfo {volume} {427}},\ \bibinfo {pages}
  {737} (\bibinfo {year} {2004})}\BibitemShut {NoStop}%
\bibitem [{\citenamefont {Milo}\ \emph {et~al.}(2002)\citenamefont {Milo},
  \citenamefont {Shen-Orr}, \citenamefont {Itzkovitz}, \citenamefont {Kashtan},
  \citenamefont {Chklovskii},\ and\ \citenamefont {Alon}}]{milo2002network}%
  \BibitemOpen
  \bibfield  {author} {\bibinfo {author} {\bibfnamefont {R.}~\bibnamefont
  {Milo}}, \bibinfo {author} {\bibfnamefont {S.}~\bibnamefont {Shen-Orr}},
  \bibinfo {author} {\bibfnamefont {S.}~\bibnamefont {Itzkovitz}}, \bibinfo
  {author} {\bibfnamefont {N.}~\bibnamefont {Kashtan}}, \bibinfo {author}
  {\bibfnamefont {D.}~\bibnamefont {Chklovskii}},\ and\ \bibinfo {author}
  {\bibfnamefont {U.}~\bibnamefont {Alon}},\ }\bibfield  {title} {\bibinfo
  {title} {Network motifs: simple building blocks of complex networks},\
  }\href@noop {} {\bibfield  {journal} {\bibinfo  {journal} {Science}\ }\textbf
  {\bibinfo {volume} {298}},\ \bibinfo {pages} {824} (\bibinfo {year}
  {2002})}\BibitemShut {NoStop}%
\bibitem [{\citenamefont {Kauffman}\ \emph {et~al.}(2003)\citenamefont
  {Kauffman}, \citenamefont {Peterson}, \citenamefont {Samuelsson},\ and\
  \citenamefont {Troein}}]{kauffman2003random}%
  \BibitemOpen
  \bibfield  {author} {\bibinfo {author} {\bibfnamefont {S.}~\bibnamefont
  {Kauffman}}, \bibinfo {author} {\bibfnamefont {C.}~\bibnamefont {Peterson}},
  \bibinfo {author} {\bibfnamefont {B.}~\bibnamefont {Samuelsson}},\ and\
  \bibinfo {author} {\bibfnamefont {C.}~\bibnamefont {Troein}},\ }\bibfield
  {title} {\bibinfo {title} {Random boolean network models and the yeast
  transcriptional network},\ }\href@noop {} {\bibfield  {journal} {\bibinfo
  {journal} {Proceedings of the National Academy of Sciences}\ }\textbf
  {\bibinfo {volume} {100}},\ \bibinfo {pages} {14796} (\bibinfo {year}
  {2003})}\BibitemShut {NoStop}%
\bibitem [{\citenamefont {Li}\ \emph {et~al.}(2013)\citenamefont {Li},
  \citenamefont {Bianco}, \citenamefont {Zhang},\ and\ \citenamefont
  {Tang}}]{li2013generic}%
  \BibitemOpen
  \bibfield  {author} {\bibinfo {author} {\bibfnamefont {Z.}~\bibnamefont
  {Li}}, \bibinfo {author} {\bibfnamefont {S.}~\bibnamefont {Bianco}}, \bibinfo
  {author} {\bibfnamefont {Z.}~\bibnamefont {Zhang}},\ and\ \bibinfo {author}
  {\bibfnamefont {C.}~\bibnamefont {Tang}},\ }\bibfield  {title} {\bibinfo
  {title} {Generic properties of random gene regulatory networks},\ }\href@noop
  {} {\bibfield  {journal} {\bibinfo  {journal} {Quantitative biology}\
  }\textbf {\bibinfo {volume} {1}},\ \bibinfo {pages} {253} (\bibinfo {year}
  {2013})}\BibitemShut {NoStop}%
\bibitem [{\citenamefont {Edwards}\ and\ \citenamefont
  {Glass}(2000)}]{edwards2000combinatorial}%
  \BibitemOpen
  \bibfield  {author} {\bibinfo {author} {\bibfnamefont {R.}~\bibnamefont
  {Edwards}}\ and\ \bibinfo {author} {\bibfnamefont {L.}~\bibnamefont
  {Glass}},\ }\bibfield  {title} {\bibinfo {title} {Combinatorial explosion in
  model gene networks},\ }\href@noop {} {\bibfield  {journal} {\bibinfo
  {journal} {Chaos: An Interdisciplinary Journal of Nonlinear Science}\
  }\textbf {\bibinfo {volume} {10}},\ \bibinfo {pages} {691} (\bibinfo {year}
  {2000})}\BibitemShut {NoStop}%
\bibitem [{\citenamefont {Stern}\ \emph {et~al.}(2014)\citenamefont {Stern},
  \citenamefont {Sompolinsky},\ and\ \citenamefont
  {Abbott}}]{stern2014dynamics}%
  \BibitemOpen
  \bibfield  {author} {\bibinfo {author} {\bibfnamefont {M.}~\bibnamefont
  {Stern}}, \bibinfo {author} {\bibfnamefont {H.}~\bibnamefont {Sompolinsky}},\
  and\ \bibinfo {author} {\bibfnamefont {L.}~\bibnamefont {Abbott}},\
  }\bibfield  {title} {\bibinfo {title} {Dynamics of random neural networks
  with bistable units},\ }\href@noop {} {\bibfield  {journal} {\bibinfo
  {journal} {Physical Review E}\ }\textbf {\bibinfo {volume} {90}},\ \bibinfo
  {pages} {062710} (\bibinfo {year} {2014})}\BibitemShut {NoStop}%
\bibitem [{\citenamefont {Stern}\ \emph {et~al.}(2007)\citenamefont {Stern},
  \citenamefont {Dror}, \citenamefont {Stolovicki}, \citenamefont {Brenner},\
  and\ \citenamefont {Braun}}]{Stern07}%
  \BibitemOpen
  \bibfield  {author} {\bibinfo {author} {\bibfnamefont {S.}~\bibnamefont
  {Stern}}, \bibinfo {author} {\bibfnamefont {T.}~\bibnamefont {Dror}},
  \bibinfo {author} {\bibfnamefont {E.}~\bibnamefont {Stolovicki}}, \bibinfo
  {author} {\bibfnamefont {N.}~\bibnamefont {Brenner}},\ and\ \bibinfo {author}
  {\bibfnamefont {E.}~\bibnamefont {Braun}},\ }\bibfield  {title} {\bibinfo
  {title} {Genome-wide transcriptional plasticity underlies cellular adaptation
  to novel challenge},\ }\href@noop {} {\bibfield  {journal} {\bibinfo
  {journal} {Molecular Systems Biology}\ }\textbf {\bibinfo {volume} {3}},\
  \bibinfo {pages} {106} (\bibinfo {year} {2007})}\BibitemShut {NoStop}%
\bibitem [{\citenamefont {Braun}(2015)}]{Braun15}%
  \BibitemOpen
  \bibfield  {author} {\bibinfo {author} {\bibfnamefont {E.}~\bibnamefont
  {Braun}},\ }\bibfield  {title} {\bibinfo {title} {The unforeseen challenge:
  from genotype-to-phenotype in cell populations},\ }\href@noop {} {\bibfield
  {journal} {\bibinfo  {journal} {Reports on Progress in Physics}\ }\textbf
  {\bibinfo {volume} {78}},\ \bibinfo {pages} {036602} (\bibinfo {year}
  {2015})}\BibitemShut {NoStop}%
\bibitem [{\citenamefont {Schreier}\ \emph {et~al.}(2017)\citenamefont
  {Schreier}, \citenamefont {Soen},\ and\ \citenamefont
  {Brenner}}]{schreier2017exploratory}%
  \BibitemOpen
  \bibfield  {author} {\bibinfo {author} {\bibfnamefont {H.~I.}\ \bibnamefont
  {Schreier}}, \bibinfo {author} {\bibfnamefont {Y.}~\bibnamefont {Soen}},\
  and\ \bibinfo {author} {\bibfnamefont {N.}~\bibnamefont {Brenner}},\
  }\bibfield  {title} {\bibinfo {title} {Exploratory adaptation in large random
  networks},\ }\href@noop {} {\bibfield  {journal} {\bibinfo  {journal} {Nature
  communications}\ }\textbf {\bibinfo {volume} {8}},\ \bibinfo {pages} {1}
  (\bibinfo {year} {2017})}\BibitemShut {NoStop}%
\bibitem [{\citenamefont {Kaplan}\ \emph {et~al.}(2021)\citenamefont {Kaplan},
  \citenamefont {Reich}, \citenamefont {Oster}, \citenamefont {Maoz},
  \citenamefont {Levin-Reisman}, \citenamefont {Ronin}, \citenamefont {Gefen},
  \citenamefont {Agam},\ and\ \citenamefont {Balaban}}]{kaplan2021observation}%
  \BibitemOpen
  \bibfield  {author} {\bibinfo {author} {\bibfnamefont {Y.}~\bibnamefont
  {Kaplan}}, \bibinfo {author} {\bibfnamefont {S.}~\bibnamefont {Reich}},
  \bibinfo {author} {\bibfnamefont {E.}~\bibnamefont {Oster}}, \bibinfo
  {author} {\bibfnamefont {S.}~\bibnamefont {Maoz}}, \bibinfo {author}
  {\bibfnamefont {I.}~\bibnamefont {Levin-Reisman}}, \bibinfo {author}
  {\bibfnamefont {I.}~\bibnamefont {Ronin}}, \bibinfo {author} {\bibfnamefont
  {O.}~\bibnamefont {Gefen}}, \bibinfo {author} {\bibfnamefont
  {O.}~\bibnamefont {Agam}},\ and\ \bibinfo {author} {\bibfnamefont {N.~Q.}\
  \bibnamefont {Balaban}},\ }\bibfield  {title} {\bibinfo {title} {Observation
  of universal ageing dynamics in antibiotic persistence},\ }\href@noop {}
  {\bibfield  {journal} {\bibinfo  {journal} {Nature}\ }\textbf {\bibinfo
  {volume} {600}},\ \bibinfo {pages} {290} (\bibinfo {year}
  {2021})}\BibitemShut {NoStop}%
\bibitem [{\citenamefont {Tosa}\ and\ \citenamefont
  {Pizer}(1971)}]{tosa1971effect}%
  \BibitemOpen
  \bibfield  {author} {\bibinfo {author} {\bibfnamefont {T.}~\bibnamefont
  {Tosa}}\ and\ \bibinfo {author} {\bibfnamefont {L.~I.}\ \bibnamefont
  {Pizer}},\ }\bibfield  {title} {\bibinfo {title} {Effect of serine
  hydroxamate on the growth of escherichia coli},\ }\href@noop {} {\bibfield
  {journal} {\bibinfo  {journal} {Journal of bacteriology}\ }\textbf {\bibinfo
  {volume} {106}},\ \bibinfo {pages} {966} (\bibinfo {year}
  {1971})}\BibitemShut {NoStop}%
\bibitem [{\citenamefont {Ritort}\ and\ \citenamefont
  {Sollich}(2003)}]{Ritort2003}%
  \BibitemOpen
  \bibfield  {author} {\bibinfo {author} {\bibfnamefont {F.}~\bibnamefont
  {Ritort}}\ and\ \bibinfo {author} {\bibfnamefont {P.}~\bibnamefont
  {Sollich}},\ }\bibfield  {title} {\bibinfo {title} {Glassy dynamics of
  kinetically constrained models},\ }\href@noop {} {\bibfield  {journal}
  {\bibinfo  {journal} {Advances in physics}\ }\textbf {\bibinfo {volume}
  {52}},\ \bibinfo {pages} {219} (\bibinfo {year} {2003})}\BibitemShut
  {NoStop}%
\bibitem [{\citenamefont {Struik}(1977)}]{Struik78}%
  \BibitemOpen
  \bibfield  {author} {\bibinfo {author} {\bibfnamefont {L.~C.~E.}\
  \bibnamefont {Struik}},\ }\bibfield  {title} {\bibinfo {title} {Physical
  aging in amorphous polymers and other materials},\ }\href@noop {} {\
  (\bibinfo {year} {1977})}\BibitemShut {NoStop}%
\bibitem [{\citenamefont {Hodge}(1995)}]{hodge1995physical}%
  \BibitemOpen
  \bibfield  {author} {\bibinfo {author} {\bibfnamefont {I.~M.}\ \bibnamefont
  {Hodge}},\ }\bibfield  {title} {\bibinfo {title} {Physical aging in polymer
  glasses},\ }\href@noop {} {\bibfield  {journal} {\bibinfo  {journal}
  {Science}\ }\textbf {\bibinfo {volume} {267}},\ \bibinfo {pages} {1945}
  (\bibinfo {year} {1995})}\BibitemShut {NoStop}%
\bibitem [{\citenamefont {Hwa}\ \emph {et~al.}(2003)\citenamefont {Hwa},
  \citenamefont {Marinari}, \citenamefont {Sneppen},\ and\ \citenamefont
  {Tang}}]{Hwa2003}%
  \BibitemOpen
  \bibfield  {author} {\bibinfo {author} {\bibfnamefont {T.}~\bibnamefont
  {Hwa}}, \bibinfo {author} {\bibfnamefont {E.}~\bibnamefont {Marinari}},
  \bibinfo {author} {\bibfnamefont {K.}~\bibnamefont {Sneppen}},\ and\ \bibinfo
  {author} {\bibfnamefont {L.-h.}\ \bibnamefont {Tang}},\ }\bibfield  {title}
  {\bibinfo {title} {Localization of denaturation bubbles in random dna
  sequences},\ }\href@noop {} {\bibfield  {journal} {\bibinfo  {journal}
  {Proceedings of the National Academy of Sciences}\ }\textbf {\bibinfo
  {volume} {100}},\ \bibinfo {pages} {4411} (\bibinfo {year}
  {2003})}\BibitemShut {NoStop}%
\bibitem [{\citenamefont {Matan}\ \emph {et~al.}(2002)\citenamefont {Matan},
  \citenamefont {Williams}, \citenamefont {Witten},\ and\ \citenamefont
  {Nagel}}]{Matan2002}%
  \BibitemOpen
  \bibfield  {author} {\bibinfo {author} {\bibfnamefont {K.}~\bibnamefont
  {Matan}}, \bibinfo {author} {\bibfnamefont {R.~B.}\ \bibnamefont {Williams}},
  \bibinfo {author} {\bibfnamefont {T.~A.}\ \bibnamefont {Witten}},\ and\
  \bibinfo {author} {\bibfnamefont {S.~R.}\ \bibnamefont {Nagel}},\ }\bibfield
  {title} {\bibinfo {title} {Crumpling a thin sheet},\ }\href@noop {}
  {\bibfield  {journal} {\bibinfo  {journal} {Physical Review Letters}\
  }\textbf {\bibinfo {volume} {88}},\ \bibinfo {pages} {076101} (\bibinfo
  {year} {2002})}\BibitemShut {NoStop}%
\bibitem [{\citenamefont {Lahini}\ \emph {et~al.}(2017)\citenamefont {Lahini},
  \citenamefont {Gottesman}, \citenamefont {Amir},\ and\ \citenamefont
  {Rubinstein}}]{Lahini2017}%
  \BibitemOpen
  \bibfield  {author} {\bibinfo {author} {\bibfnamefont {Y.}~\bibnamefont
  {Lahini}}, \bibinfo {author} {\bibfnamefont {O.}~\bibnamefont {Gottesman}},
  \bibinfo {author} {\bibfnamefont {A.}~\bibnamefont {Amir}},\ and\ \bibinfo
  {author} {\bibfnamefont {S.~M.}\ \bibnamefont {Rubinstein}},\ }\bibfield
  {title} {\bibinfo {title} {Nonmonotonic aging and memory retention in
  disordered mechanical systems},\ }\href@noop {} {\bibfield  {journal}
  {\bibinfo  {journal} {Physical review letters}\ }\textbf {\bibinfo {volume}
  {118}},\ \bibinfo {pages} {085501} (\bibinfo {year} {2017})}\BibitemShut
  {NoStop}%
\bibitem [{\citenamefont {Ghofraniha}\ \emph
  {et~al.}(2007{\natexlab{a}})\citenamefont {Ghofraniha}, \citenamefont
  {Conti},\ and\ \citenamefont {Ruocco}}]{Ghofraniha2007a}%
  \BibitemOpen
  \bibfield  {author} {\bibinfo {author} {\bibfnamefont {N.}~\bibnamefont
  {Ghofraniha}}, \bibinfo {author} {\bibfnamefont {C.}~\bibnamefont {Conti}},\
  and\ \bibinfo {author} {\bibfnamefont {G.}~\bibnamefont {Ruocco}},\
  }\bibfield  {title} {\bibinfo {title} {Aging of the nonlinear optical
  susceptibility in doped colloidal suspensions},\ }\href@noop {} {\bibfield
  {journal} {\bibinfo  {journal} {Physical Review B}\ }\textbf {\bibinfo
  {volume} {75}},\ \bibinfo {pages} {224203} (\bibinfo {year}
  {2007}{\natexlab{a}})}\BibitemShut {NoStop}%
\bibitem [{\citenamefont {Ghofraniha}\ \emph
  {et~al.}(2007{\natexlab{b}})\citenamefont {Ghofraniha}, \citenamefont
  {Conti}, \citenamefont {Di~Leonardo}, \citenamefont {Ruzicka},\ and\
  \citenamefont {Ruocco}}]{Ghofraniha2007b}%
  \BibitemOpen
  \bibfield  {author} {\bibinfo {author} {\bibfnamefont {N.}~\bibnamefont
  {Ghofraniha}}, \bibinfo {author} {\bibfnamefont {C.}~\bibnamefont {Conti}},
  \bibinfo {author} {\bibfnamefont {R.}~\bibnamefont {Di~Leonardo}}, \bibinfo
  {author} {\bibfnamefont {B.}~\bibnamefont {Ruzicka}},\ and\ \bibinfo {author}
  {\bibfnamefont {G.}~\bibnamefont {Ruocco}},\ }\bibfield  {title} {\bibinfo
  {title} {Ageing of the nonlinear optical susceptibility in soft matter},\
  }\href@noop {} {\bibfield  {journal} {\bibinfo  {journal} {Journal of
  Physics: Condensed Matter}\ }\textbf {\bibinfo {volume} {19}},\ \bibinfo
  {pages} {205129} (\bibinfo {year} {2007}{\natexlab{b}})}\BibitemShut
  {NoStop}%
\bibitem [{\citenamefont {M{\'e}zard}\ \emph {et~al.}(1987)\citenamefont
  {M{\'e}zard}, \citenamefont {Parisi},\ and\ \citenamefont
  {Virasoro}}]{GlassBook}%
  \BibitemOpen
  \bibfield  {author} {\bibinfo {author} {\bibfnamefont {M.}~\bibnamefont
  {M{\'e}zard}}, \bibinfo {author} {\bibfnamefont {G.}~\bibnamefont {Parisi}},\
  and\ \bibinfo {author} {\bibfnamefont {M.~A.}\ \bibnamefont {Virasoro}},\
  }\href@noop {} {\emph {\bibinfo {title} {Spin glass theory and beyond: An
  Introduction to the Replica Method and Its Applications}}},\ Vol.~\bibinfo
  {volume} {9}\ (\bibinfo  {publisher} {World Scientific Publishing Company},\
  \bibinfo {year} {1987})\BibitemShut {NoStop}%
\bibitem [{\citenamefont {Lundgren}\ \emph {et~al.}(1983)\citenamefont
  {Lundgren}, \citenamefont {Svedlindh}, \citenamefont {Nordblad},\ and\
  \citenamefont {Beckman}}]{Lundgren83}%
  \BibitemOpen
  \bibfield  {author} {\bibinfo {author} {\bibfnamefont {L.}~\bibnamefont
  {Lundgren}}, \bibinfo {author} {\bibfnamefont {P.}~\bibnamefont {Svedlindh}},
  \bibinfo {author} {\bibfnamefont {P.}~\bibnamefont {Nordblad}},\ and\
  \bibinfo {author} {\bibfnamefont {O.}~\bibnamefont {Beckman}},\ }\bibfield
  {title} {\bibinfo {title} {Dynamics of the relaxation-time spectrum in a cumn
  spin-glass},\ }\href@noop {} {\bibfield  {journal} {\bibinfo  {journal}
  {Physical review letters}\ }\textbf {\bibinfo {volume} {51}},\ \bibinfo
  {pages} {911} (\bibinfo {year} {1983})}\BibitemShut {NoStop}%
\bibitem [{\citenamefont {Cugliandolo}\ and\ \citenamefont
  {Kurchan}(1993)}]{cugliandolo1993analytical}%
  \BibitemOpen
  \bibfield  {author} {\bibinfo {author} {\bibfnamefont {L.~F.}\ \bibnamefont
  {Cugliandolo}}\ and\ \bibinfo {author} {\bibfnamefont {J.}~\bibnamefont
  {Kurchan}},\ }\bibfield  {title} {\bibinfo {title} {Analytical solution of
  the off-equilibrium dynamics of a long-range spin-glass model},\ }\href@noop
  {} {\bibfield  {journal} {\bibinfo  {journal} {Physical Review Letters}\
  }\textbf {\bibinfo {volume} {71}},\ \bibinfo {pages} {173} (\bibinfo {year}
  {1993})}\BibitemShut {NoStop}%
\bibitem [{\citenamefont {Vincent}\ \emph {et~al.}(1997)\citenamefont
  {Vincent}, \citenamefont {Hammann}, \citenamefont {Ocio}, \citenamefont
  {Bouchaud},\ and\ \citenamefont {Cugliandolo}}]{vincent1997slow}%
  \BibitemOpen
  \bibfield  {author} {\bibinfo {author} {\bibfnamefont {E.}~\bibnamefont
  {Vincent}}, \bibinfo {author} {\bibfnamefont {J.}~\bibnamefont {Hammann}},
  \bibinfo {author} {\bibfnamefont {M.}~\bibnamefont {Ocio}}, \bibinfo {author}
  {\bibfnamefont {J.-P.}\ \bibnamefont {Bouchaud}},\ and\ \bibinfo {author}
  {\bibfnamefont {L.~F.}\ \bibnamefont {Cugliandolo}},\ }\bibfield  {title}
  {\bibinfo {title} {Slow dynamics and aging in spin glasses},\ }in\ \href@noop
  {} {\emph {\bibinfo {booktitle} {Complex Behaviour of Glassy Systems}}}\
  (\bibinfo  {publisher} {Springer},\ \bibinfo {year} {1997})\ pp.\ \bibinfo
  {pages} {184--219}\BibitemShut {NoStop}%
\bibitem [{\citenamefont {Debenedetti}\ and\ \citenamefont
  {Stillinger}(2001)}]{Debenedetti2001}%
  \BibitemOpen
  \bibfield  {author} {\bibinfo {author} {\bibfnamefont {P.~G.}\ \bibnamefont
  {Debenedetti}}\ and\ \bibinfo {author} {\bibfnamefont {F.~H.}\ \bibnamefont
  {Stillinger}},\ }\bibfield  {title} {\bibinfo {title} {Supercooled liquids
  and the glass transition},\ }\href@noop {} {\bibfield  {journal} {\bibinfo
  {journal} {Nature}\ }\textbf {\bibinfo {volume} {410}},\ \bibinfo {pages}
  {259} (\bibinfo {year} {2001})}\BibitemShut {NoStop}%
\bibitem [{\citenamefont {Vaknin}\ \emph {et~al.}(2001)\citenamefont {Vaknin},
  \citenamefont {Pollak},\ and\ \citenamefont {Ovadyahu}}]{Vaknin2001}%
  \BibitemOpen
  \bibfield  {author} {\bibinfo {author} {\bibfnamefont {A.}~\bibnamefont
  {Vaknin}}, \bibinfo {author} {\bibfnamefont {M.}~\bibnamefont {Pollak}},\
  and\ \bibinfo {author} {\bibfnamefont {Z.}~\bibnamefont {Ovadyahu}},\
  }\bibfield  {title} {\bibinfo {title} {Memory and aging in an electron
  glass},\ }\href@noop {} {\bibfield  {journal} {\bibinfo  {journal} {SPRINGER
  PROCEEDINGS IN PHYSICS}\ }\textbf {\bibinfo {volume} {87}},\ \bibinfo {pages}
  {995} (\bibinfo {year} {2001})}\BibitemShut {NoStop}%
\bibitem [{\citenamefont {Mehta}(2004)}]{MehtaBook}%
  \BibitemOpen
  \bibfield  {author} {\bibinfo {author} {\bibfnamefont {M.~L.}\ \bibnamefont
  {Mehta}},\ }\href@noop {} {\emph {\bibinfo {title} {Random matrices}}}\
  (\bibinfo  {publisher} {Elsevier},\ \bibinfo {year} {2004})\BibitemShut
  {NoStop}%
\bibitem [{\citenamefont {Glauber}(1963)}]{glauber1963time}%
  \BibitemOpen
  \bibfield  {author} {\bibinfo {author} {\bibfnamefont {R.~J.}\ \bibnamefont
  {Glauber}},\ }\bibfield  {title} {\bibinfo {title} {Time-dependent statistics
  of the ising model},\ }\href@noop {} {\bibfield  {journal} {\bibinfo
  {journal} {Journal of mathematical physics}\ }\textbf {\bibinfo {volume}
  {4}},\ \bibinfo {pages} {294} (\bibinfo {year} {1963})}\BibitemShut {NoStop}%
\bibitem [{\citenamefont {Sherrington}\ and\ \citenamefont
  {Kirkpatrick}(1975)}]{SK}%
  \BibitemOpen
  \bibfield  {author} {\bibinfo {author} {\bibfnamefont {D.}~\bibnamefont
  {Sherrington}}\ and\ \bibinfo {author} {\bibfnamefont {S.}~\bibnamefont
  {Kirkpatrick}},\ }\bibfield  {title} {\bibinfo {title} {Solvable model of a
  spin-glass},\ }\href@noop {} {\bibfield  {journal} {\bibinfo  {journal}
  {Physical review letters}\ }\textbf {\bibinfo {volume} {35}},\ \bibinfo
  {pages} {1792} (\bibinfo {year} {1975})}\BibitemShut {NoStop}%
\bibitem [{\citenamefont {Palmer}(1982)}]{Palmer82}%
  \BibitemOpen
  \bibfield  {author} {\bibinfo {author} {\bibfnamefont {R.}~\bibnamefont
  {Palmer}},\ }\bibfield  {title} {\bibinfo {title} {Broken ergodicity},\
  }\href@noop {} {\bibfield  {journal} {\bibinfo  {journal} {Advances in
  Physics}\ }\textbf {\bibinfo {volume} {31}},\ \bibinfo {pages} {669}
  (\bibinfo {year} {1982})}\BibitemShut {NoStop}%
\bibitem [{\citenamefont {Kinzel}(1986)}]{Kinzel86}%
  \BibitemOpen
  \bibfield  {author} {\bibinfo {author} {\bibfnamefont {W.}~\bibnamefont
  {Kinzel}},\ }\bibfield  {title} {\bibinfo {title} {Remanent magnetization of
  the infinite-range ising spin glass},\ }\href@noop {} {\bibfield  {journal}
  {\bibinfo  {journal} {Physical Review B}\ }\textbf {\bibinfo {volume} {33}},\
  \bibinfo {pages} {5086} (\bibinfo {year} {1986})}\BibitemShut {NoStop}%
\bibitem [{\citenamefont {Henkel}\ and\ \citenamefont
  {Kinzel}(1987)}]{Kinzel87}%
  \BibitemOpen
  \bibfield  {author} {\bibinfo {author} {\bibfnamefont {R.}~\bibnamefont
  {Henkel}}\ and\ \bibinfo {author} {\bibfnamefont {W.}~\bibnamefont
  {Kinzel}},\ }\bibfield  {title} {\bibinfo {title} {Metastable states of the
  sk model of spin glasses},\ }\href@noop {} {\bibfield  {journal} {\bibinfo
  {journal} {Journal of Physics A: Mathematical and General}\ }\textbf
  {\bibinfo {volume} {20}},\ \bibinfo {pages} {L727} (\bibinfo {year}
  {1987})}\BibitemShut {NoStop}%
\bibitem [{\citenamefont {Sibani}\ and\ \citenamefont
  {Hoffmann}(1989)}]{Sibani89}%
  \BibitemOpen
  \bibfield  {author} {\bibinfo {author} {\bibfnamefont {P.}~\bibnamefont
  {Sibani}}\ and\ \bibinfo {author} {\bibfnamefont {K.~H.}\ \bibnamefont
  {Hoffmann}},\ }\bibfield  {title} {\bibinfo {title} {Hierarchical models for
  aging and relaxation of spin glasses},\ }\href@noop {} {\bibfield  {journal}
  {\bibinfo  {journal} {Physical review letters}\ }\textbf {\bibinfo {volume}
  {63}},\ \bibinfo {pages} {2853} (\bibinfo {year} {1989})}\BibitemShut
  {NoStop}%
\bibitem [{\citenamefont {Kohring}\ and\ \citenamefont
  {Schreckenberg}(1991)}]{Kohring91}%
  \BibitemOpen
  \bibfield  {author} {\bibinfo {author} {\bibfnamefont {G.}~\bibnamefont
  {Kohring}}\ and\ \bibinfo {author} {\bibfnamefont {M.}~\bibnamefont
  {Schreckenberg}},\ }\bibfield  {title} {\bibinfo {title} {Numerical studies
  of the spin-flip dynamics in the sk-model},\ }\href@noop {} {\bibfield
  {journal} {\bibinfo  {journal} {Journal de Physique I}\ }\textbf {\bibinfo
  {volume} {1}},\ \bibinfo {pages} {1087} (\bibinfo {year} {1991})}\BibitemShut
  {NoStop}%
\bibitem [{\citenamefont {Bouchaud}(1992)}]{Bouchaud92}%
  \BibitemOpen
  \bibfield  {author} {\bibinfo {author} {\bibfnamefont {J.-P.}\ \bibnamefont
  {Bouchaud}},\ }\bibfield  {title} {\bibinfo {title} {Weak ergodicity breaking
  and aging in disordered systems},\ }\href@noop {} {\bibfield  {journal}
  {\bibinfo  {journal} {Journal de Physique I}\ }\textbf {\bibinfo {volume}
  {2}},\ \bibinfo {pages} {1705} (\bibinfo {year} {1992})}\BibitemShut
  {NoStop}%
\bibitem [{\citenamefont {Parisi}\ and\ \citenamefont
  {Ritort}(1993)}]{Parisi93}%
  \BibitemOpen
  \bibfield  {author} {\bibinfo {author} {\bibfnamefont {G.}~\bibnamefont
  {Parisi}}\ and\ \bibinfo {author} {\bibfnamefont {F.}~\bibnamefont
  {Ritort}},\ }\bibfield  {title} {\bibinfo {title} {The remanent magnetization
  in spin-glass models},\ }\href@noop {} {\bibfield  {journal} {\bibinfo
  {journal} {Journal de Physique I}\ }\textbf {\bibinfo {volume} {3}},\
  \bibinfo {pages} {969} (\bibinfo {year} {1993})}\BibitemShut {NoStop}%
\bibitem [{\citenamefont {Cugliandolo}\ and\ \citenamefont
  {Kurchan}(1994)}]{Kurchan94a}%
  \BibitemOpen
  \bibfield  {author} {\bibinfo {author} {\bibfnamefont {L.~F.}\ \bibnamefont
  {Cugliandolo}}\ and\ \bibinfo {author} {\bibfnamefont {J.}~\bibnamefont
  {Kurchan}},\ }\bibfield  {title} {\bibinfo {title} {On the out-of-equilibrium
  relaxation of the sherrington-kirkpatrick model},\ }\href@noop {} {\bibfield
  {journal} {\bibinfo  {journal} {Journal of Physics A: Mathematical and
  General}\ }\textbf {\bibinfo {volume} {27}},\ \bibinfo {pages} {5749}
  (\bibinfo {year} {1994})}\BibitemShut {NoStop}%
\bibitem [{\citenamefont {Cugliandolo}\ \emph {et~al.}(1994)\citenamefont
  {Cugliandolo}, \citenamefont {Kurchan},\ and\ \citenamefont
  {Ritort}}]{Kurchan94b}%
  \BibitemOpen
  \bibfield  {author} {\bibinfo {author} {\bibfnamefont {L.}~\bibnamefont
  {Cugliandolo}}, \bibinfo {author} {\bibfnamefont {J.}~\bibnamefont
  {Kurchan}},\ and\ \bibinfo {author} {\bibfnamefont {F.}~\bibnamefont
  {Ritort}},\ }\bibfield  {title} {\bibinfo {title} {Evidence of aging in
  spin-glass mean-field models},\ }\href@noop {} {\bibfield  {journal}
  {\bibinfo  {journal} {Physical Review B}\ }\textbf {\bibinfo {volume} {49}},\
  \bibinfo {pages} {6331} (\bibinfo {year} {1994})}\BibitemShut {NoStop}%
\bibitem [{\citenamefont {Scharnagl}\ \emph {et~al.}(1995)\citenamefont
  {Scharnagl}, \citenamefont {Opper},\ and\ \citenamefont {Kinzel}}]{Kinzel95}%
  \BibitemOpen
  \bibfield  {author} {\bibinfo {author} {\bibfnamefont {A.}~\bibnamefont
  {Scharnagl}}, \bibinfo {author} {\bibfnamefont {M.}~\bibnamefont {Opper}},\
  and\ \bibinfo {author} {\bibfnamefont {W.}~\bibnamefont {Kinzel}},\
  }\bibfield  {title} {\bibinfo {title} {On the relaxation of infinite-range
  spin glasses},\ }\href@noop {} {\bibfield  {journal} {\bibinfo  {journal}
  {Journal of Physics A: Mathematical and General}\ }\textbf {\bibinfo {volume}
  {28}},\ \bibinfo {pages} {5721} (\bibinfo {year} {1995})}\BibitemShut
  {NoStop}%
\bibitem [{\citenamefont {Yoshino}\ \emph {et~al.}(1997)\citenamefont
  {Yoshino}, \citenamefont {Hukushima},\ and\ \citenamefont
  {Takayama}}]{Yoshino97}%
  \BibitemOpen
  \bibfield  {author} {\bibinfo {author} {\bibfnamefont {H.}~\bibnamefont
  {Yoshino}}, \bibinfo {author} {\bibfnamefont {K.}~\bibnamefont {Hukushima}},\
  and\ \bibinfo {author} {\bibfnamefont {H.}~\bibnamefont {Takayama}},\
  }\bibfield  {title} {\bibinfo {title} {Relaxational modes and aging in the
  glauber dynamics of the sherrington-kirkpatrick model},\ }\href@noop {}
  {\bibfield  {journal} {\bibinfo  {journal} {Progress of Theoretical Physics
  Supplement}\ }\textbf {\bibinfo {volume} {126}},\ \bibinfo {pages} {107}
  (\bibinfo {year} {1997})}\BibitemShut {NoStop}%
\bibitem [{\citenamefont {Eissfeller}\ and\ \citenamefont
  {Opper}(1994)}]{Opper94}%
  \BibitemOpen
  \bibfield  {author} {\bibinfo {author} {\bibfnamefont {H.}~\bibnamefont
  {Eissfeller}}\ and\ \bibinfo {author} {\bibfnamefont {M.}~\bibnamefont
  {Opper}},\ }\bibfield  {title} {\bibinfo {title} {Mean-field monte carlo
  approach to the sherrington-kirkpatrick model with asymmetric couplings},\
  }\href@noop {} {\bibfield  {journal} {\bibinfo  {journal} {Physical Review
  E}\ }\textbf {\bibinfo {volume} {50}},\ \bibinfo {pages} {709} (\bibinfo
  {year} {1994})}\BibitemShut {NoStop}%
\bibitem [{\citenamefont {Bastolla}\ and\ \citenamefont
  {Parisi}(1998)}]{Parisi98}%
  \BibitemOpen
  \bibfield  {author} {\bibinfo {author} {\bibfnamefont {U.}~\bibnamefont
  {Bastolla}}\ and\ \bibinfo {author} {\bibfnamefont {G.}~\bibnamefont
  {Parisi}},\ }\bibfield  {title} {\bibinfo {title} {Relevant elements,
  magnetization and dynamical properties in kauffman networks: A numerical
  study},\ }\href@noop {} {\bibfield  {journal} {\bibinfo  {journal} {Physica
  D: Nonlinear Phenomena}\ }\textbf {\bibinfo {volume} {115}},\ \bibinfo
  {pages} {203} (\bibinfo {year} {1998})}\BibitemShut {NoStop}%
\bibitem [{\citenamefont {Nagar}\ \emph {et~al.}(2021)\citenamefont {Nagar},
  \citenamefont {Ecker}, \citenamefont {Loewenthal}, \citenamefont {Avram},
  \citenamefont {Ben-Meir}, \citenamefont {Biran}, \citenamefont {Ron},\ and\
  \citenamefont {Pupko}}]{nagar2021harnessing}%
  \BibitemOpen
  \bibfield  {author} {\bibinfo {author} {\bibfnamefont {N.}~\bibnamefont
  {Nagar}}, \bibinfo {author} {\bibfnamefont {N.}~\bibnamefont {Ecker}},
  \bibinfo {author} {\bibfnamefont {G.}~\bibnamefont {Loewenthal}}, \bibinfo
  {author} {\bibfnamefont {O.}~\bibnamefont {Avram}}, \bibinfo {author}
  {\bibfnamefont {D.}~\bibnamefont {Ben-Meir}}, \bibinfo {author}
  {\bibfnamefont {D.}~\bibnamefont {Biran}}, \bibinfo {author} {\bibfnamefont
  {E.}~\bibnamefont {Ron}},\ and\ \bibinfo {author} {\bibfnamefont
  {T.}~\bibnamefont {Pupko}},\ }\bibfield  {title} {\bibinfo {title}
  {Harnessing machine learning to unravel protein degradation in escherichia
  coli},\ }\href@noop {} {\bibfield  {journal} {\bibinfo  {journal} {Msystems}\
  }\textbf {\bibinfo {volume} {6}},\ \bibinfo {pages} {e01296} (\bibinfo {year}
  {2021})}\BibitemShut {NoStop}%
\bibitem [{\citenamefont {Uhlenbeck}\ and\ \citenamefont
  {Ornstein}(1930)}]{Uhlenbeck1930}%
  \BibitemOpen
  \bibfield  {author} {\bibinfo {author} {\bibfnamefont {G.~E.}\ \bibnamefont
  {Uhlenbeck}}\ and\ \bibinfo {author} {\bibfnamefont {L.~S.}\ \bibnamefont
  {Ornstein}},\ }\bibfield  {title} {\bibinfo {title} {On the theory of the
  brownian motion},\ }\href@noop {} {\bibfield  {journal} {\bibinfo  {journal}
  {Physical review}\ }\textbf {\bibinfo {volume} {36}},\ \bibinfo {pages} {823}
  (\bibinfo {year} {1930})}\BibitemShut {NoStop}%
\bibitem [{\citenamefont {Nyberg}\ \emph {et~al.}(2016)\citenamefont {Nyberg},
  \citenamefont {Ambj{\"o}rnsson},\ and\ \citenamefont {Lizana}}]{Nyberg16}%
  \BibitemOpen
  \bibfield  {author} {\bibinfo {author} {\bibfnamefont {M.}~\bibnamefont
  {Nyberg}}, \bibinfo {author} {\bibfnamefont {T.}~\bibnamefont
  {Ambj{\"o}rnsson}},\ and\ \bibinfo {author} {\bibfnamefont {L.}~\bibnamefont
  {Lizana}},\ }\bibfield  {title} {\bibinfo {title} {A simple method to
  calculate first-passage time densities with arbitrary initial conditions},\
  }\href@noop {} {\bibfield  {journal} {\bibinfo  {journal} {New Journal of
  Physics}\ }\textbf {\bibinfo {volume} {18}},\ \bibinfo {pages} {063019}
  (\bibinfo {year} {2016})}\BibitemShut {NoStop}%
\end{thebibliography}%

\end{document}